\documentclass{article}

\PassOptionsToPackage{numbers, compress}{natbib}

\usepackage[preprint]{neurips_2024}

\makeatletter
\renewcommand*{\@fnsymbol}[1]{\ensuremath{\ifcase#1\or \else\@ctrerr\fi}}
\makeatother

\usepackage[utf8]{inputenc} 
\usepackage[T1]{fontenc}    
\usepackage{hyperref}       
\usepackage{url}            
\usepackage{booktabs}       
\usepackage{amsfonts}       
\usepackage{nicefrac}       
\usepackage{microtype}      
\usepackage{amsmath}
\usepackage{amsthm}
\usepackage{amssymb}
\usepackage{physics}
\usepackage[ruled,vlined]{algorithm2e}
\usepackage{multirow}
\usepackage{graphicx}
\usepackage[belowskip=-6pt]{caption}
\usepackage{graphicx}
\usepackage{amsmath}
\usepackage{subcaption}

\usepackage{graphicx}
\usepackage{amssymb}
\usepackage{multirow}

\makeatletter

\usepackage{xcolor}

\newcommand {\ignore}[1]{}

\usepackage{lineno}
\usepackage{amsmath}
\usepackage{subcaption}

\title{Unraveling particle dark matter with \\ Physics-Informed Neural Networks}

\author{%
  M.~P. Bento, \ H.~B. Câmara, \  J.~F. Seabra \\
  Departamento de F\'{\i}sica and CFTP\\
  Instituto Superior T\'ecnico, Universidade de Lisboa,\\
  Av. Rovisco Pais 1, 1049-001 Lisboa, Portugal \\
  \texttt{miguel.pedra.bento@tecnico.ulisboa.pt}, \\  \texttt{henrique.b.camara@tecnico.ulisboa.pt}, \\ \texttt{joao.f.seabra@tecnico.ulisboa.pt} \\
}

\begin{document}

\maketitle

\begin{abstract}

We parametrically solve the Boltzmann equations (BEs) governing freeze-in dark matter~(DM) in alternative cosmologies with Physics-Informed Neural Networks~(PINNs), a mesh-free method. Through \textit{inverse} PINNs, using a single DM experimental point -- observed relic density -- we determine the physical attributes of the theory, namely power-law cosmologies, inspired by braneworld scenarios, and particle interaction cross sections. The expansion of the Universe in such alternative cosmologies has been parameterized through a switch-like function reproducing the Hubble law at later times. Without loss of generality, we model more realistically this transition with a smooth function. We predict a distinct pair-wise relationship between power-law exponent and particle interactions: for a given cosmology with negative (positive) exponent, smaller (larger) cross sections are required to reproduce the data. Lastly, via Bayesian methods, we quantify the epistemic uncertainty of theoretical parameters found in \textit{inverse} problems.

\end{abstract}

\maketitle
\noindent

\section{Introduction}
\label{sec:intro}

The Standard Model~(SM) of particle physics, which describes the fundamental particles and their interactions, stands among the greatest achievements in theoretical physics. The predictions of the SM have been in remarkable agreement with experimental results. Namely, its last missing piece, the Higgs boson, was discovered in 2012, by ATLAS and CMS, at the Large Hadron Collider~(LHC)~\cite{ATLAS:2015yey}. Nonetheless, there are significant experimental evidences for new physics. One of the most striking examples of such new physics is the existence of dark matter~(DM), established through indirect astrophysical measurements and cosmological observations (for a recent review see Ref.~\cite{Cirelli:2024ssz}). These indicate that about $27\%$ of the total energy in the Universe is DM, being about $5$ times more abundant than ordinary baryonic matter~\cite{Planck:2018vyg}. Precisely, the observed cold DM~(CDM) relic abundance obtained by the Planck satellite data is $\Omega_{\text{CDM}} h^2 = 0.120 \pm 0.001$~\cite{Planck:2018vyg}. This data point assumes the $\Lambda$CDM model, which describes the evolution of our Universe, its matter content, structure formation and accelerated expansion following the Hubble law, based on the Big-Bang theory. A viable particle DM candidate must be electrically neutral, non-baryonic, stable and non-relativistic (cold). Since no such particle is found in the SM, we must explore beyond the SM extensions. 

Among the numerous particle DM proposals stand out weakly interacting massive particles (WIMPs)~\cite{Kolb:1988aj}. WIMPs undergo thermal freeze-out after initially being in thermal equilibrium with the SM particle bath (visible sector). In the simplest scenario, their relic abundance is produced via annihilation and inverse annihilation processes -- $\text{DM} \ \text{DM} \leftrightarrow \text{SM} \ \text{SM}$ -- which constitute the thermally-averaged cross-section. The evolution of particle DM yield is governed by Boltzmann equations~(BEs), that include the effects of the Hubble expansion of the Universe and particle interactions. WIMPs account for the observed CDM abundance typically with masses around the electroweak scale and interact with the visible sector through couplings of the order of weak interactions. This so-called "WIMP miracle" is being tested at direct detection experiments, such as LUX-ZEPLIN~(LZ)~\cite{LZ:2022lsv} and PandaX-4T~\cite{PandaX-4T:2021bab}. Currently, the experimental exclusion limits constrain the WIMP parameter space almost until the "neutrino floor"~\cite{Billard:2013qya}. The lack of a direct experimental detection challenges the WIMP paradigm, pushing theorists to explore other DM generation mechanisms. A viable alternative is the generation of the observed DM abundance through freeze-in~\cite{Hall:2009bx} (for a review see Ref.~\cite{Bernal:2017kxu}). In this scenario the DM candidate couples feebly to the visible sector, never entering in thermal equilibrium. For this reason, it is referred as feebly interacting massive particle (FIMP)~\cite{Hall:2009bx}. Freeze-in production mechanisms have been extensively studied in the literature for example through the scalar portal~\cite{Yaguna:2011qn} (see Ref.~\cite{Chu:2011be} for an instructive overview of DM portals), with a dark gauge boson~\cite{Feldman:2006wd,Kang:2010mh,Chu:2013jja}, as well as via non-renormalizable operators leading to the so-called ultraviolet (UV) freeze-in scenarios~\cite{Choi:2005vq,Hall:2009bx,Yaguna:2011ei,Krauss:2013wfa,Elahi:2014fsa,Roland:2014vba,Roland:2015yoa,McDonald:2015ljz,Barman:2020plp}. Moreover, from a theoretical perspective, the developments in string theory lead to the study of branes in extra-dimensional scenarios, opening alternative cosmological frameworks which have attracted significant attention in the context of DM phenomenology, see e.g. the recent Refs.~\cite{Maity:2018dgy,Garcia:2020eof,Garcia:2020wiy,Garcia:2021gsy,Giudice:2000ex,Barman:2022tzk,Bhattiprolu:2022sdd,Harigaya:2014waa,Harigaya:2019tzu,Okada:2021uqk,Ghosh:2022fws,Haque:2021mab,Ahmed:2022tfm,Bernal:2022wck,Bernal:2023ura,Bernal:2018kcw,Bernal:2019mhf,Bhatia:2020itt,Hamdan:2017psw,Arcadi:2024wwg,Barman:2024tjt,Gonzalez:2024dtb,Gonzalez:2024rhs,Silva-Malpartida:2024emu,Bernal:2024yhu,Barman:2024mqo,Cosme:2024ndc,Cosme:2023xpa,Haque:2023awl,Haque:2024zdq,Mondal:2025awq}. Notably, braneworld cosmologies based on the Randall-Sundrum~(RS)~\cite{Randall:1999vf} model, such as RS II, have been extensively studied in the literature~\cite{Langlois:2002bb,Brax:2003fv,Brax:2004xh}. Additionally, the extension of the RS II model with the Gauss-Bonnet~(GB) invariant leads to a distinct cosmology~\cite{Charmousis:2002rc,Maeda:2003vq}. In both cases, the Standard Friedmann equations are altered, modifying the cosmological evolution of our Universe. In fact, it was found that these scenarios lead to specific power-law cosmologies in the early Universe with the Hubble law being reproduced at low temperatures~\cite{Okada:2004nc,Okada:2009xe}. This transition can be parameterized via a switch-like function in the BEs governing WIMP~\cite{Okada:2004nc,Okada:2009xe,Liu:2023qhv} or FIMP~\cite{Bernal:2018kcw,Baules:2019zwk} particle abundance. Naturally, the relic density will depend on this modified early Universe expansion law. In order to comprehensively study the phenomenology of WIMP and FIMP DM models, state-of-the-art numerical tools are available, with the prominent ones being \texttt{MicrOmegas}~\cite{Belanger:2014vza,Belanger:2018ccd} and \texttt{DarkSUSY}~\cite{Bringmann:2018lay}. These rely on finite element methods~(FEMs) to solve the BEs, taking into account the couplings and masses of concrete model realizations. A plethora of testable experimental observables are computed with these tools, e.g. relic density, spin independent/dependent interaction cross sections, etc. With this in mind, we explore machine learning~(ML) techniques that can help unravel DM models.

The impact of ML has been so significant that the 2024 physics Nobel prize recognized the application of physics concepts to its foundations. In particular, deep learning techniques based on neural networks~(NNs) have been very successful, solving a multitude of problems in different fields such as computer vision, natural language processing, game theory, science and engineering, among others. The application of various ML algorithms has also made its way into particle physics with an ever growing popularity~\cite{hepmllivingreview}. Typically, these are incorporated into experiments, which generally require the analysis of large volumes of high-dimensional and highly-correlated data. But even from a theoretical standpoint, particle physics provides problems where ML has been shown to be useful -- see e.g. Refs.~\cite{Craven:2021ems, Matchev:2023mii, Wojcik:2023usm, Kawai:2024pws, Cheung:2024svk,Nishimura:2020nre,Nishimura:2024apb,Nishimura:2025knz}. In this work we will make use of an algorithm called Physics-Informed Neural Networks~(PINNs) (for a review see Ref.~\cite{toscano2024pinns}). Recent developments have sought to extend the performance of the basic PINN architecture, e.g.~\cite{zhao2024pinnsformertransformerbasedframeworkphysicsinformed,Wang_2025,seiler2025stifftransferlearningphysicsinformed}, by addressing the so-called
failure modes of PINNs~\cite{wang2020understandingmitigatinggradientpathologies,basir2022criticalinvestigationfailuremodes}. The idea of the basic PINN is to make use of NNs as universal approximators~\cite{HORNIK1989359}, where prior knowledge of physics is fed into the learning process of the NNs. The aim is to solve partial and/or ordinary differential equations~(PDEs/ODEs)~\cite{RAISSI2019686}. The NN is able to achieve such task via the minimization of a loss function encoding the residuals of the given PDEs or ODEs, as well as initial and/or final and/or boundary conditions. Adding this information to the loss function optimizes the learning process of a classical NN. In fact, PINNs do not require large amounts of training data and can make use of automatic differentiation~(AD)~\cite{baydin2018automaticdifferentiationmachinelearning} to evaluate all differential operators entering such equations. PINNs have been used to tackle a multiplicity of problems, e.g. involving stiff equations~\cite{stiff_pinns,Ji_2021,baty2023solvingstiffordinarydifferential,baty2023solvingdifferentialequationsusing,seiler2025stifftransferlearningphysicsinformed}; heat equations~\cite{Zobeiry_2021}; Navier-Stokes equations for fluid dynamics~\cite{naderibeni2024learningsolutionsparametricnavierstokes,PhysRevLett.130.244002}; modeling astrophysical objects with data and equations of state~\cite{BATY2023100734,10.1093/mnras/stad3320,Andika:2024gsc,auddy2023grinnphysicsinformedneuralnetwork,Chen:2022gok,Stefanou:2023jxk,livermore2024reconstructionsjupitersmagneticfield,dahlbüdding2024approximatingrayleighscatteringexoplanetary}; studying neutrino flavor conversion in supernova~\cite{Abbar:2023ltx,Abbar:2024ynh}; modeling DM halo~\cite{Dai:2023kip}; using gravitational lensing data for DM mapping~\cite{Ojha:2024LensPINN}; obtaining quasi-normal modes of black holes in General Relativity~\cite{Cornell:2022enn,Lobos:2024fzj,Patel:2024iij}; study of cosmological equations that govern the background dynamics of the Universe~\cite{Chantada:2022bdf}; renormalization group flows on the lattice~\cite{Yokota:2023czk} and only very recently in the context of high-energy-physics~\cite{Vatellis:2024vjl,Terin:2024iyy,Ihssen:2024ihp,Kou:2025qsg}. One type of generic problem PINNs have been used for are the so-called \textit{forward} problems. Namely, given a physical law, usually characterized by PDEs or ODEs, and initial/boundary conditions, a solution is found by the PINN, corresponding to a physical quantity. But there is another interesting class of problems that can be solved by PINNs: \textit{inverse} problems. In this case, the observable quantities are given to the PINN which subsequently determines the physical laws and/or parameters of a given physical theory that can satisfy such observables. This is of particular interest, since PINNs can potentially be a tool to discover models and/or a given models' parameter space that inherently satisfy experimental observables.

Inspired by the above ideas, in this work, we solve the paradigmatic freeze-in DM BE in alternative cosmologies using PINNs. Our goal is to showcase how PINNs are a powerful technology for exploring theoretical physics models motivated by experimental data in the context of particle DM -- an approach that has so far not been explored in the literature. Specifically, we aim to address the following:
\begin{itemize}
    \item \textit{\textit{Forward} problem:} Given a physical theory, what experimentally testable predictions does it lead to? In our case, we will give to the PINN the information about particle interactions and cosmological models. The PINN will compute the predictions for relic density of each specific theoretical scenario, to be confronted with the observed CDM relic density value.
    
    \item \textit{\textit{Inverse} problem:} Given the experimentally observed data, what are the physical theories that can explain this data? Specifically, we will give to the PINN the information about the observed CDM abundance. The PINN will find the particle interactions and/or cosmological models that account for this singular experimental data point. 
\end{itemize}
The physics background is presented in Section~\ref{sec:physics}. We identify the FIMP DM BE in non-standard cosmology. We parameterize alternative cosmologies through a transition switch-like function between power-law and Hubble expansion. Next, in Section~\ref{sec:numerical}, we recast the BE in a suitable form for numerical resolution. In Section~\ref{sec:forward}, the \textit{forward} problem is solved by the PINN parametrically as a function of the interaction cross section for the Standard, RS and GB cosmologies. The \textit{inverse} problems are addressed in Section~\ref{sec:inverse}. Namely, in Section~\ref{sec:inverseC}, for the aforementioned cosmological cases, the PINN determines the interaction cross section strength necessary to reproduce the observed CDM abundance. In Section~\ref{sec:inversegamma}, we make use of the PINN to discover other power-law cosmological scenarios that explain the data. Furthermore, in Section~\ref{sec:inversetransition}, we test the PINN for a proposed modification to the transition function that parameterizes the alternative cosmologies. We study, in Section~\ref{sec:bayesian}, a Bayesian method, in order to quantify the epistemic uncertainty of the \textit{inverse} PINN in determining model parameters. Finally, our concluding remarks are drawn in Section~\ref{sec:concl}.

\section{Freeze-in dark matter in alternative cosmologies}
\label{sec:physics}

We consider a DM candidate that in the primordial Universe never reached thermal equilibrium with its surrounding SM particle bath, with initial abundance zero or negligibly small. We assume there are no sizeable DM self-interactions, with the DM relic abundance produced through freeze-in via annihilation processes -- $\text{SM} \ \text{SM} \rightarrow \text{DM} \ \text{DM}$. This DM candidate is commonly referred as FIMP~\cite{Hall:2009bx}. The evolution of the FIMP DM number density is governed by the following BE,
\begin{equation}
\frac{d Y(x)}{d x} = \frac{1}{x^2} \frac{S(m)}{H(m)} \langle \sigma v \rangle  Y_{\text{eq}}(x)^2 \; , \; Y(x_0 \ll 1)=Y_0 \ll 1 \; ,
\label{eq:freezein1}
\end{equation}
written in terms of the particle number density per comoving volume $Y \equiv n / S$, also known as yield, and $x \equiv m/T$, with $m$ being the mass of our DM candidate and $T$ the temperature of the Universe. The total thermally averaged cross section $\langle \sigma v \rangle$ encodes the production processes -- $\text{SM} \ \text{SM} \rightarrow \text{DM} \ \text{DM}$. The initial abundance $Y_0$ is negligibly small at very early times $x_0$. 

In Standard Big Bang cosmology, the Hubble parameter, which governs the expansion rate of the Universe, and entropy density are given by
\begin{equation}
H(T) = \sqrt{\frac{4 \pi^3}{45}} g_\ast^{1/2} \frac{T^2}{m_{\text{Pl}}} \; , \; S(T) = \frac{2 \pi^2}{45} g_{\ast S} T^3 \; .
\label{eq:Hands}
\end{equation}
Here, $m_{\text{Pl}} = 1.22 \times 10^{19}$ GeV is the Planck mass, whereas $g_\ast$ and $g_{\ast S}$ correspond to the effective number of relativistic degrees of freedom related to the energy and entropy density, respectively. The latter are usually taken to be $g_\ast \simeq g_{\ast S} \simeq 100$. The Friedmann equation are modified in a non-standard cosmological history, resulting in a different expansion law of the Universe. In this work, we will consider power-law cosmology and parameterize the modified Hubble parameter through a switch-like function~\cite{Okada:2004nc,Okada:2009xe,Liu:2023qhv},
\begin{align}
    H(x) \rightarrow H(x) \times F(x,x_t,\gamma) \; , \;
    F(x,x_t,\gamma) =\left\{  
  \begin{array}{ll}  \displaystyle
    \left(\frac{x_t}{x}\right)^\gamma & \ \text{for} \ x < x_t \vspace{0.3cm} \\
    \displaystyle \hspace{0.3cm} 1 \hspace{0.5cm} & \ \text{for} \ x > x_t
  \end{array} \right. \; ,
  \label{eq:Hpowerlaw}
\end{align}
where, via $x_t = m/T_t$, we introduce a transition temperature $T_t$, at which the modified expansion law approaches the Standard picture, and $\gamma$ is the power-law parameter. $F(x,x_t,\gamma)$ stands as an assumption that captures the transition between power and Hubble law. As it turns out, it is a great approximation for the string inspired extra-dimension frameworks of GB and RS braneworld cosmologies where $\gamma = -2/3$ and $\gamma = 2$, respectively. Evidently, a complete numerical resolution of the modified Friedman equations in these scenarios would yield the full result.

The equilibrium value for the DM particle number density per comoving volume $Y_{\text{eq}}$ is,
\begin{equation}
    Y_{\text{eq}}(x) = \frac{45}{4 \pi^4} \frac{g}{g_{\ast S}} x^2 \mathcal{K}_2(x) \xrightarrow[x \gg 3]{} \frac{45}{\sqrt{32 \pi^7}} \frac{g}{g_{\ast S}} x^{3/2} \exp(-x) \; ,
    \label{eq:Yeq}
\end{equation}
with $\mathcal{K}_n$ being the modified Bessel function of second kind of order $n$ and $g$ is the number of degrees of freedom of our DM candidate. Here we consider, without loss of generality, DM to be a scalar particle, so $g=1$. In the above we show the asymptotic limit $x \gg 3$ of $Y_{\text{eq}}$, useful for the case under consideration of cold relics.

The solution of the BE~\eqref{eq:freezein1} provides $Y(x)$, from which by taking the limit $x \rightarrow \infty$, we obtain the relic density value at present time,
\begin{equation}
    \Omega h^2 = \frac{S_0}{\rho_{\text{crit}}^0/h^2} m Y(x \rightarrow \infty) \; ,
    \label{eq:relic}
\end{equation}
with $S_0 = 2.89 \times 10^3$ cm$^{-3}$ being the entropy at present time, $\rho_{\text{crit}}^0 = 1.05 h^2 \times 10^{-5}$ GeV$/$cm$^{3}$ the present critical energy density and $h = H_0 / (100$ km$/$s$/$Mpc) the reduced Hubble constant~\cite{ParticleDataGroup:2022pth}. The above needs to be confronted with the experimentally observed value $\Omega_{\text{CDM}} h^2 = 0.120 \pm 0.001$~\cite{Planck:2018vyg}.

There are many theoretical physics models one can envisage and our objective in this work is to unravel different theories for freeze-in DM based on the physical parameters $\langle \sigma v \rangle$, $x_t$ and $\gamma$. The thermally averaged cross section $\langle \sigma v \rangle$ encodes the interaction between our FIMP DM candidate and the visible sector of a given model. The power-law cosmology is characterized by the transition temperature reflected in $x_t$ and the $\gamma$ exponent. Among all the possible scenarios, we focus on a benchmark where $\langle \sigma v \rangle$ is constant. We have:
\begin{itemize}

    \item \textit{Standard cosmology --} This is the simplest scenario, where we work with the Standard Hubble law. In such case the BE~\eqref{eq:freezein1} leads to an exact solution for the DM particle yield,
    \begin{align}
    Y_{\text{Std}}(x) & = \frac{135 5^{1/2}}{128 \ \pi^{13/2}} \frac{g^2 g_{\ast S}}{g_{\ast}^{5/2}} m_{\text{Pl}} \ m  \langle \sigma v \rangle \ \left[1 - \exp(-2 x) (1 - 2 x) \right] \; .
    \end{align}
    By taking the $x \rightarrow \infty$ limit, we obtain the relic density [see Eq.~\eqref{eq:relic}],
    \begin{align}
    \Omega h^2_{\text{Std}} & = \frac{S_0}{\rho_{\text{crit}}^0/h^2} \frac{135 5^{1/2}}{128 \ \pi^{13/2}} \frac{g^2 g_{\ast S}}{g_{\ast}^{5/2}} m_{\text{Pl}} \ m^2  \langle \sigma v \rangle \; .
    \label{eq:Oh2Std}
    \end{align}
    Notice that $\Omega h^2_{\text{Std}}$ depends linearly on the particle interaction strength $\langle \sigma v \rangle$. This is the typical behavior for FIMP DM. As an example, we fix $g_\ast \simeq g_{\ast S} \simeq 100$ and $g=1$. Requiring that $\Omega h^2_{\text{Std}}$ accounts for the entire CDM budget, i.e. $\Omega h^2_{\text{Std}} = \Omega_{\text{CDM}} h^2 = 0.12$, then $\langle \sigma v \rangle \in 2.58 \times [10^{-25},10^{-29}] \ \text{GeV}^{-2}$ for DM masses $m \in [10,10^{3}] \ \text{GeV}$. We remark that the FIMP DM interactions are extremely "feeble" compared to the electroweak scale cross sections probed at the LHC, with typical strength around $\mathcal{O}(10^{-9}) \ \text{GeV}^{-2}$.

    \item \textit{Alternative cosmologies --} The Hubble parameter in~\eqref{eq:Hpowerlaw} modifies the Standard FIMP DM BE~\eqref{eq:freezein1}. By considering $\langle \sigma v \rangle$ constant, we can exactly solve the modified BE for $T>T_t$, i.e. if we ignore the transition between power and Hubble law. We obtain,
    \begin{align}
    Y_{(x_t,\gamma)}(x) & = \frac{135 5^{1/2}}{128 \ \pi^{13/2}} \frac{g^2 g_{\ast S}}{g_{\ast}^{5/2}} m_{\text{Pl}} \ m  \langle \sigma v \rangle \ \frac{\left[\Gamma(2 + \gamma) - 
   \Gamma(2 + \gamma, 2 x) \right]}{(2 x_t)^\gamma} \; ,
   \label{eq:Ycosmo}
    \end{align}
    with the yield now depending on the $(x_t,\gamma)$ parameters. Furthermore, the upper incomplete gamma function is defined as
    \begin{equation}
        \Gamma(a,b) = \int_{b}^{\infty} dy \ y^{a-1} \exp(-y) \; , \; a,b \in \mathbb{C} \; ,
    \end{equation}
    where for $\text{Re} \ a >0$, the integral converges absolutely and is related to the usual Euler gamma function through $\Gamma(a)=\Gamma(a,0)$. Once again, taking $x \rightarrow \infty$ yields
    \begin{equation}
    \Omega h^2_{(x_t,\gamma)} = \Omega h^2_{\text{Std}} \frac{\Gamma(2 + \gamma)}{(2 x_t)^\gamma} \; ,
    \label{eq:Oh2approxNS}
    \end{equation}
    with $\Omega h^2_{\text{Std}}$ being given by Eq.~\eqref{eq:Oh2Std}. Thus, the power-law cosmology has the effect of rescaling the Standard relic density by a factor $\Gamma(2 + \gamma)/(2 x_t)^\gamma$. In fact, taking $x_t>1$ for $\gamma>0$ ($\gamma<0$), power-law cosmology leads to a reduced (enhanced) $\Omega h^2_{\text{Std}}$ value. We wish to reinforce the fact that the above derivation only holds for the case $x_t>1$. Therefore, in order to capture the full effect of the modified cosmological law of the Universe, parameterized via Eq.~\eqref{eq:Hpowerlaw}, one needs to solve the BE numerically.
    
\end{itemize}
%

\section{Recasting the Boltzmann equation for freeze-in dark matter}
\label{sec:numerical}

The BE identified in the previous section must be solved across a wide range of the variable $x$ over which the yield spans several orders of magnitude. Then, it is judicious to rewrite Eq.~\eqref{eq:freezein1} in a more tractable form by first making the change of variables
\begin{align}
z \equiv \ln x \; , \; W(z) \equiv \ln Y(z) \; .
\label{eq:variablesnum}
\end{align}
The performance of the PINN can be further improved if we normalize $W(z)$ and $z$. Since the derivative $dY(x)/dx$ is always positive in Eq.~\eqref{eq:freezein1} and the logarithm is a strictly monotonically increasing function, any initial condition (IC) we impose establishes the minimum of $W(z)$. Considering $z_0>0$ the point where such minimum occurs with $W_0 \equiv W(z_0)$, and denoting the final point as $z_f$, we rescale $W(z)$ and $z$ as follows,
\begin{equation}
    W_n(z) = - W(z) / W_0 \; , \; z_n = z/ z_f \; .
    \label{eq:Wn}
\end{equation}
The minimum of $W_n(z)$ is $-1$, while its maximum is within the interval $[-1,0[$.

Next we define [see Eq.~\eqref{eq:Hands}],
\begin{align}
    \exp[C(m, \langle \sigma v \rangle)] \equiv\frac{\langle \sigma v \rangle S(m)}{H(m)} = \sqrt{\frac{\pi}{45}} \frac{g_{\ast S}}{g_\ast^{1/2}} \ m_{\text{Pl}} \ m  \ \langle \sigma v \rangle \, ,
    \label{eq:Cdef}
\end{align}
which encodes the information about the interactions among DM and ordinary matter via $\langle \sigma v \rangle$. Furthermore, introducing the rectified linear unit (ReLU) function $\text{ReLU}(x) \equiv \max(0,x)$ -- commonly used as an activation function in NNs -- and taking $z_t = \ln x_t$, we have [see Eq.~\eqref{eq:Hpowerlaw}]
\begin{equation}
    F(z, z_t, \gamma) = \exp\left[\gamma \ \text{ReLU}(z_t-z)\right] \; .
    \label{eq:Fnew}
\end{equation}
The above function parameterizes the cosmological history models of our Universe. In Section~\ref{sec:inversetransition}, we will propose a novel "smooth" and more realistic way to parameterize this function. For the equilibrium particle yield we take the asymptotic form of Eq.~\eqref{eq:Yeq}, resulting in
\begin{align}
W_{\text{eq}}(z) = \ln (\frac{45}{\sqrt{32 \pi^7}} \frac{g}{g_{\ast s}}) + \frac{3}{2} z - \exp(z) \; .
\label{eq:Weq}
\end{align}
Finally, the BE for FIMP DM of Eq.~\eqref{eq:freezein1}, in alternative cosmologies, is now written as
\begin{equation}
\mathcal{E}\left[z, W, \frac{d W}{dz} ; C, z_t, \gamma\right] \equiv \frac{d W}{dz} - \exp(C - \gamma \ \text{ReLU}(z_t-z) - z + 2 W_{\text{eq}} - W) = 0 \; ,
\label{eq:freezein2}
\end{equation}
with an IC equal to $W_0$ defined at some point $z_0$. The values we choose for these quantities are presented in the upcoming sections. Before proceeding, we should emphasize that the standard cosmological scenario is obtained by setting $\gamma \rightarrow 0$.

As a conventional numerical tool to solve the above BE, we use the initial value ODE solver \texttt{solve\textunderscore ivp} from the open-source \texttt{SciPy} \texttt{Python} library. We select the backward differentiation~(BDF) method, which is more stable for our ODE, even though it is not explicitly stiff. In what follows, this conventional FEM numerical solution will serve as a benchmark for comparison with the solutions obtained through PINNs, providing an essential cross-check of our results.

\section{\textit{Forward} problem: parametric PINN solution}
\label{sec:forward}

\begin{figure*}[t!]
        \centering
        \includegraphics[scale=0.095]{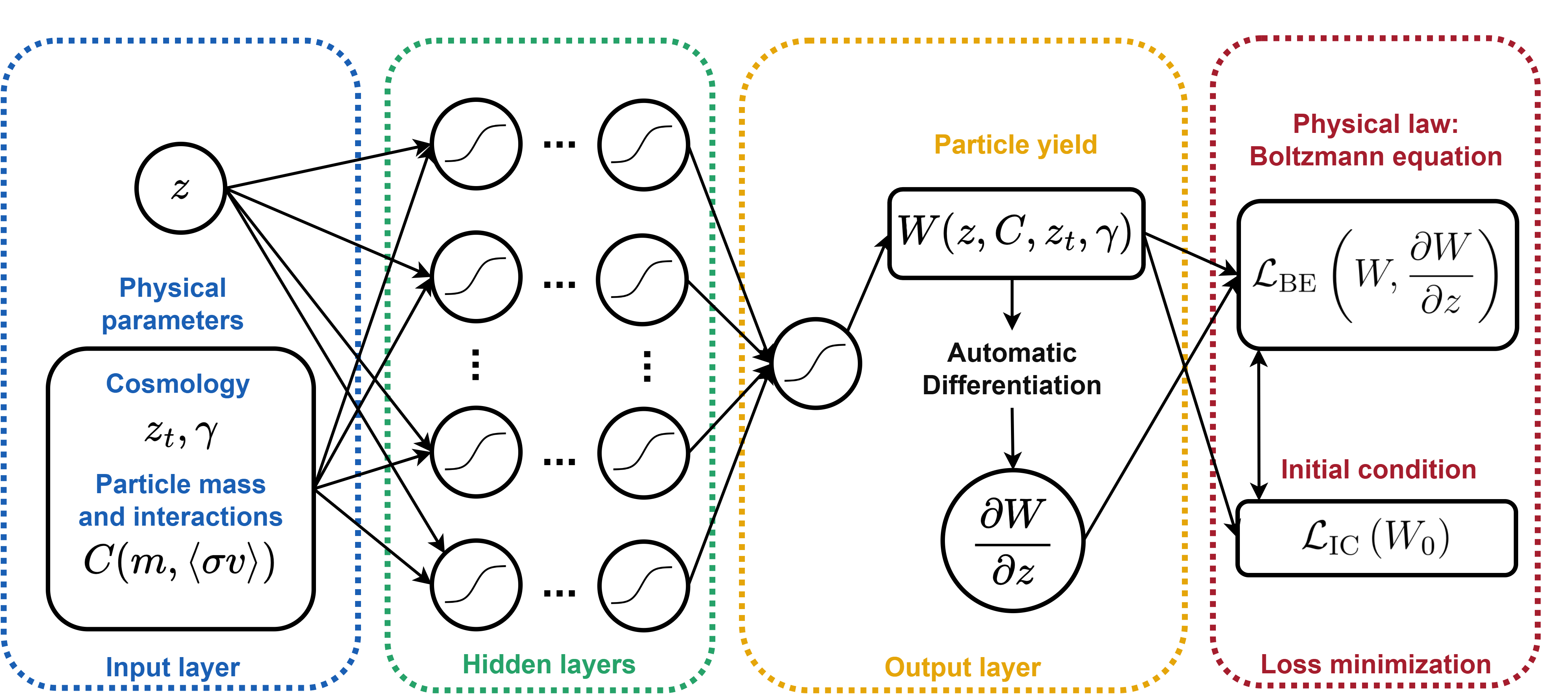}
        \caption{Schematic representation of the PINN structure for tackling the \textit{forward} problem in modeling the BE for freeze-in DM particle yield in alternative cosmology (see text for details).}
    \label{fig:forward_PINN}
    \end{figure*}

Recall that the \textit{forward} problem refers to the case where we know a theoretical physics model and/or law, and we wish to determine the observable predictions. In our case, given the particle interaction strength encoded in $C(m,\langle \sigma v \rangle)$ and cosmology parameterized via $z_t$ and $\gamma$, we want to determine the evolution of the DM particle yield and the prediction for the DM particle abundance observed in the Universe today. This evidently needs to be confronted with current experimental data on DM relic abundance. We intend to solve this \textit{forward} problem using the PINN approach, whose structure is schematically depicted in Fig.~\ref{fig:forward_PINN}. 

Concretely, we aim to solve parametrically the BE~\eqref{eq:freezein2} in Standard ($\gamma = 0$), GB ($\gamma = -2/3$) and RS ($\gamma = 2$) cosmologies for any value of $C(m,\langle\sigma v\rangle)$ within some interval. To this end, we generate 800 collocation points (the input data for our PINN) in the 2-dimensional plane formed by the variables $z$ and $C$. These are placed equidistant from each other along the direction of $z$ and within $z \in [\ln 10^{-14},\ln 100]$. On the other hand, the values of $C$ are randomly generated inside the interval $C \in [-14,-9]$, corresponding to cross-section values of $\langle \sigma v \rangle \in [2.58 \times 10^{-28}, 3.83 \times 10^{-26}] \ \text{GeV}^{-2}$ [see Eq.~\eqref{eq:Cdef}] when the DM mass is $m=100$~GeV. Without loss of generality, we use the latter DM mass value throughout the rest of this work. Note that the different procedures to choose $z$ and $C$ for each collocation point prevent the PINN from seeing those two variables as being correlated. Such behaviour would make the PINN fail to generalize its predictions to other regions of the $(z, C)$ space. Together with $\gamma$, $z$ and $C$, the parameter $z_t$ completes the set of input parameters of the \textit{forward} PINN, which we fix at $z_t = \ln 10$. 

Next to the input layer, our \textit{forward} PINN contains four dense hidden layers with 30 units each, followed by an output layer with a single unit. The activation function applied to the units of the hidden layers is the Gaussian Error Linear Unit (GELU). It is worth noting that in the context of PINNs, learnable activation functions have been successfully applied, accelerating the algorithm's convergence~\cite{Jagtap_2020}. However, we found no improvements from implementing such activation functions to the hidden layers of our PINNs, leading us to pick a simpler one. As for the activation of the output layer's unit, we apply the negative sigmoid function, forcing all the predicted values of $W_n$ to fall between $-1$ and~$0$ [see Eq.~\eqref{eq:Wn}]. All PINNs used in the rest of our work have the same exact architecture as the one described here.

    \begin{figure*}[t!]
        \centering
        \includegraphics[scale=0.3]{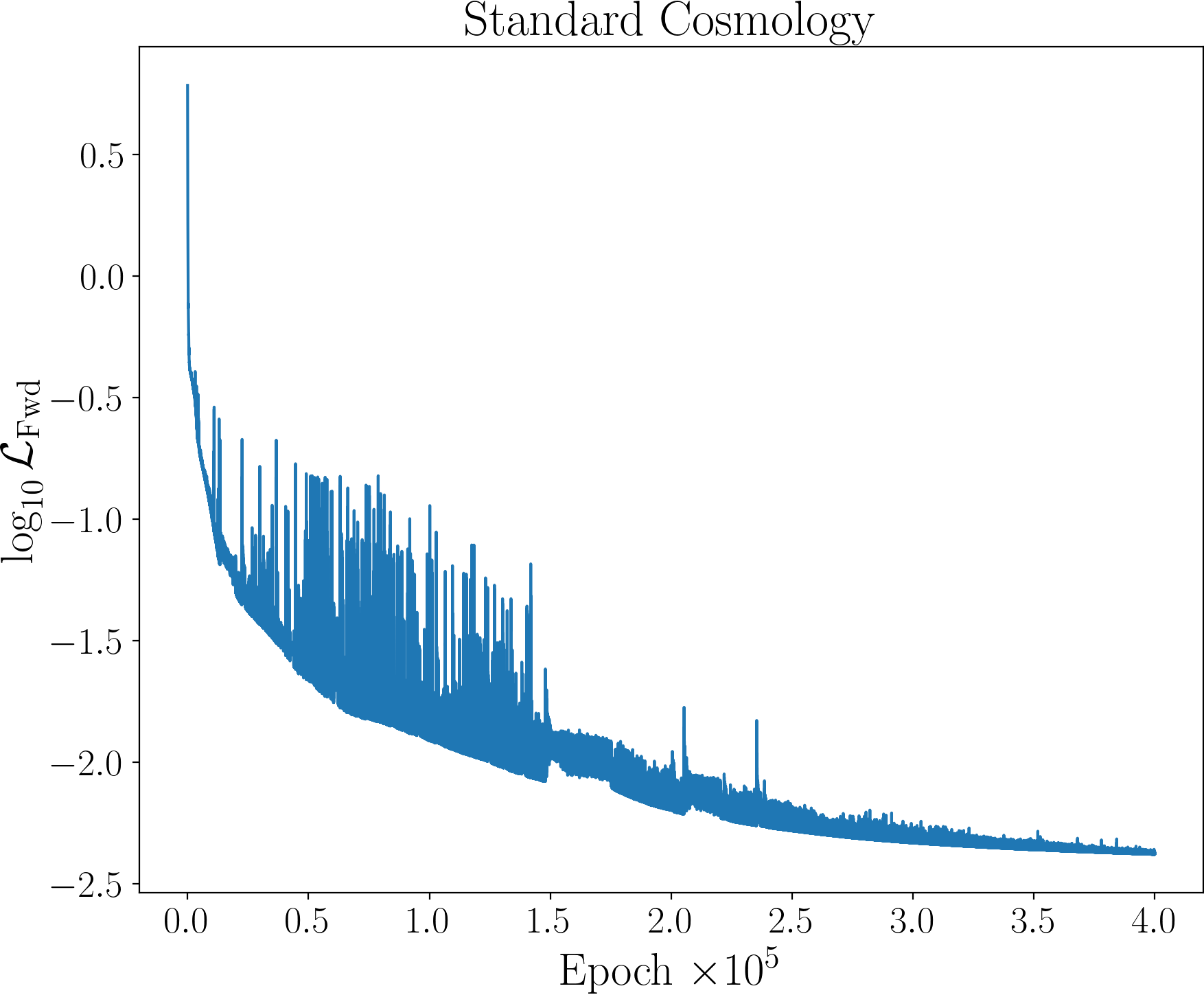}  \\ \includegraphics[scale=0.3]{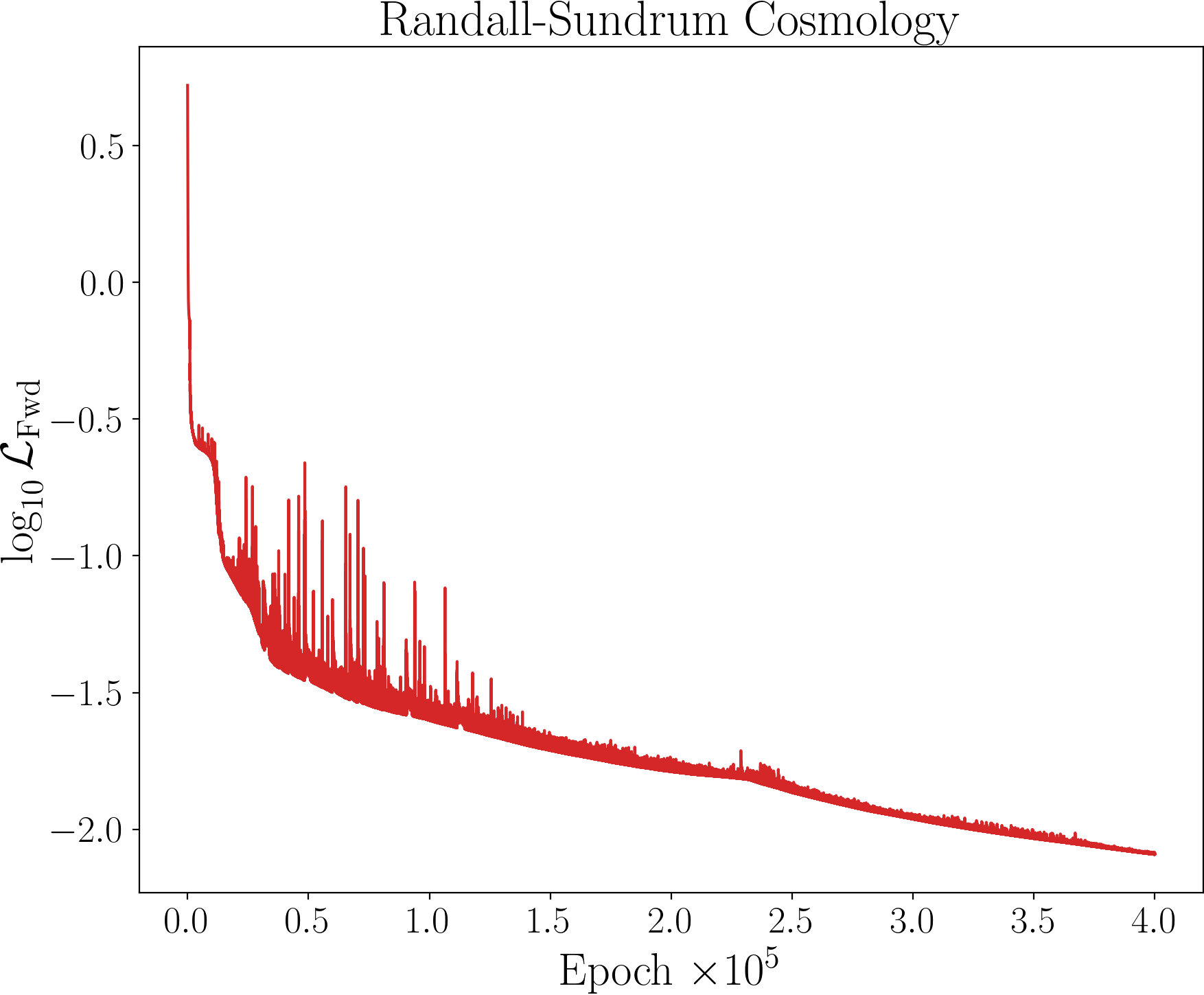} \hspace{+0.1cm} \includegraphics[scale=0.3]{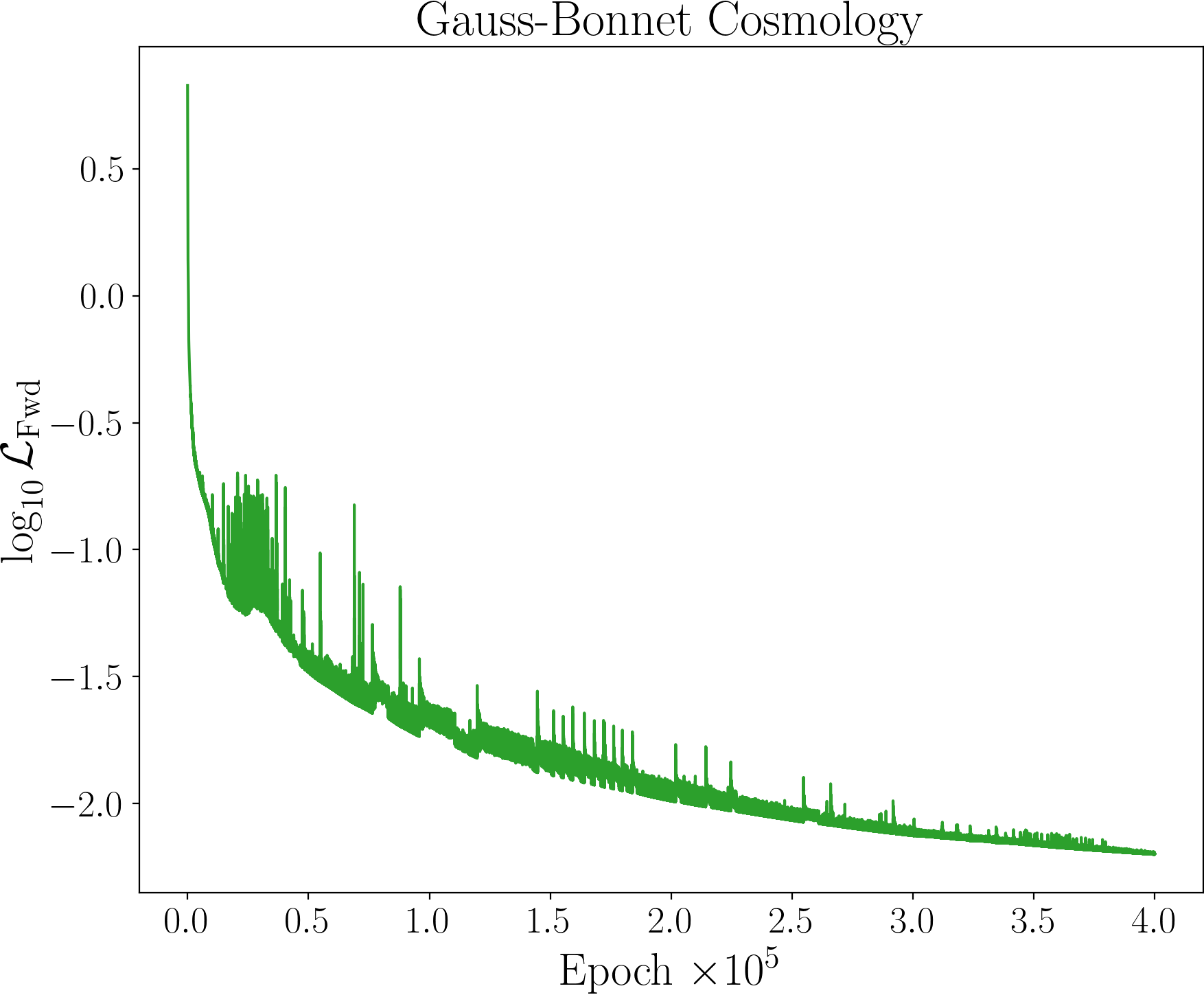}
        \caption{Evolution of the total \textit{forward} problem loss function $\mathcal{L}_{\text{Fwd}}$ [see Eqs.~\eqref{eq:Lfwd},~\eqref{eq:Lbe} and~\eqref{eq:Lic}], for the cases of Standard (top), RS (bottom left) and GB (bottom right) cosmology.}
    \label{fig:parametricforward_losses}
    \end{figure*}

The loss function to be minimized by the PINN described above is defined as follows~\footnote{Another interesting way to solve the BE is to consider $\mathcal{E}' = \left| \frac{dW}{dz} \right| - C  + z - 2W_\text{eq} + W$ and $\frac{d W}{dz} > 0$ as an extra term in the loss function,
\begin{align}
    \mathcal{L}^\prime_{\text{Fwd}} = \mathcal{L}_{\text{Fwd}} + \lambda_{+} \; \mathcal{L}_{+} \; , \;
    \mathcal{L}_{+} = \frac{1}{N_z} \sum_{j=1}^{N_z} \text{ReLU}\left[ - \frac{d W}{dz}(z_j)\right] \; , \nonumber
\end{align}
where $\lambda_{+}$ is the associated weight. Albeit a possibility, it comes at a cost, since we checked that the extra term above hinders the performance of our PINN.}:
\begin{equation}
    \mathcal{L}_{\text{Fwd}} = \lambda_{\text{BE}} \; \mathcal{L}_{\text{BE}} + \lambda_{\text{IC}} \; \mathcal{L}_{\text{IC}} \; ,
    \label{eq:Lfwd}
\end{equation}
where $\lambda_{\text{BE}}$ and $\lambda_{\text{IC}}$ denotes the weights associated with $\mathcal{L}_{\text{BE}}$ and $\mathcal{L}_{\text{IC}}$, respectively. $\mathcal{L}_{\text{BE}}$ takes the form of a mean absolute error (MAE), ensuring that the BE is satisfied at the collocation points sent through the PINN. It is given by
\begin{equation}
    \mathcal{L}_{\text{BE}} = \frac{1}{N_z} \sum_{j=1}^{N_z} \left| \mathcal{E}\left[z_j, W(z_j), \frac{d W}{dz}(z_j) ; C, z_t, \gamma\right] \right| \; ,
    \label{eq:Lbe}
\end{equation}
with $N_z=800$ being the total number of collocation points. At the final collocation point, we define $z_f \equiv z_{N_z}$. The loss term $\mathcal{L}_{\text{IC}}$ imposes the IC at the first collocation point $z_0=\ln 10^{-14}$, 
\begin{equation}
    \mathcal{L}_{\text{IC}} = \left|W(z_0)-W_0]\right| \; ,
    \label{eq:Lic}
\end{equation}
where we fix $W_0 = \ln 10^{-25}$. In order to improve the performance of PINNs in solving PDEs, some mechanisms to make loss weights adaptive have been put forward, such as learning rate annealing~\cite{wang2020understandingmitigatinggradientpathologies,wang2023expertsguidetrainingphysicsinformed} and soft attention~\cite{McClenny_2023}. Although the BE imposes similar challenges to those tackled in the aforementioned references, we found no advantages in implementing these mechanisms. Hence, $\lambda_\text{BE}$ and $\lambda_\text{IC}$, were kept constant throughout training. Note that a full hyperparameter optimization is beyond the scope of this work, which aims to demonstrate the potential of PINNs as a valuable tool for particle DM studies. Thus, for our purposes, we select the weights and the learning rate via manual tuning. We found better results when $\lambda_\text{IC}$ was fixed at a higher value than $\lambda_\text{BE}$, so we take $\lambda_\text{BE}=1$ and $\lambda_\text{IC}=10$.

The trainable parameters of the PINN are tuned by running batch gradient descent for $4\times10^5$ epochs using the Adam optimizer. The learning rate $\eta$, initially equal to $10^{-4}$, is updated every 1000 epochs by using a decay rate of $0.99$. The importance of setting this exponential decay of the learning rate is reflected by the plots in Fig.~\ref{fig:parametricforward_losses}, which show the evolution of the loss function for each cosmology throughout training. In all three cases, the oscillations of $\mathcal{L}_{\text{Fwd}}$ observed in early epochs become much smaller as the learning rate decreases, helping the loss function converge steadily to its minimum. 

    \begin{figure*}[t!]
        \centering
        \includegraphics[scale=0.35]{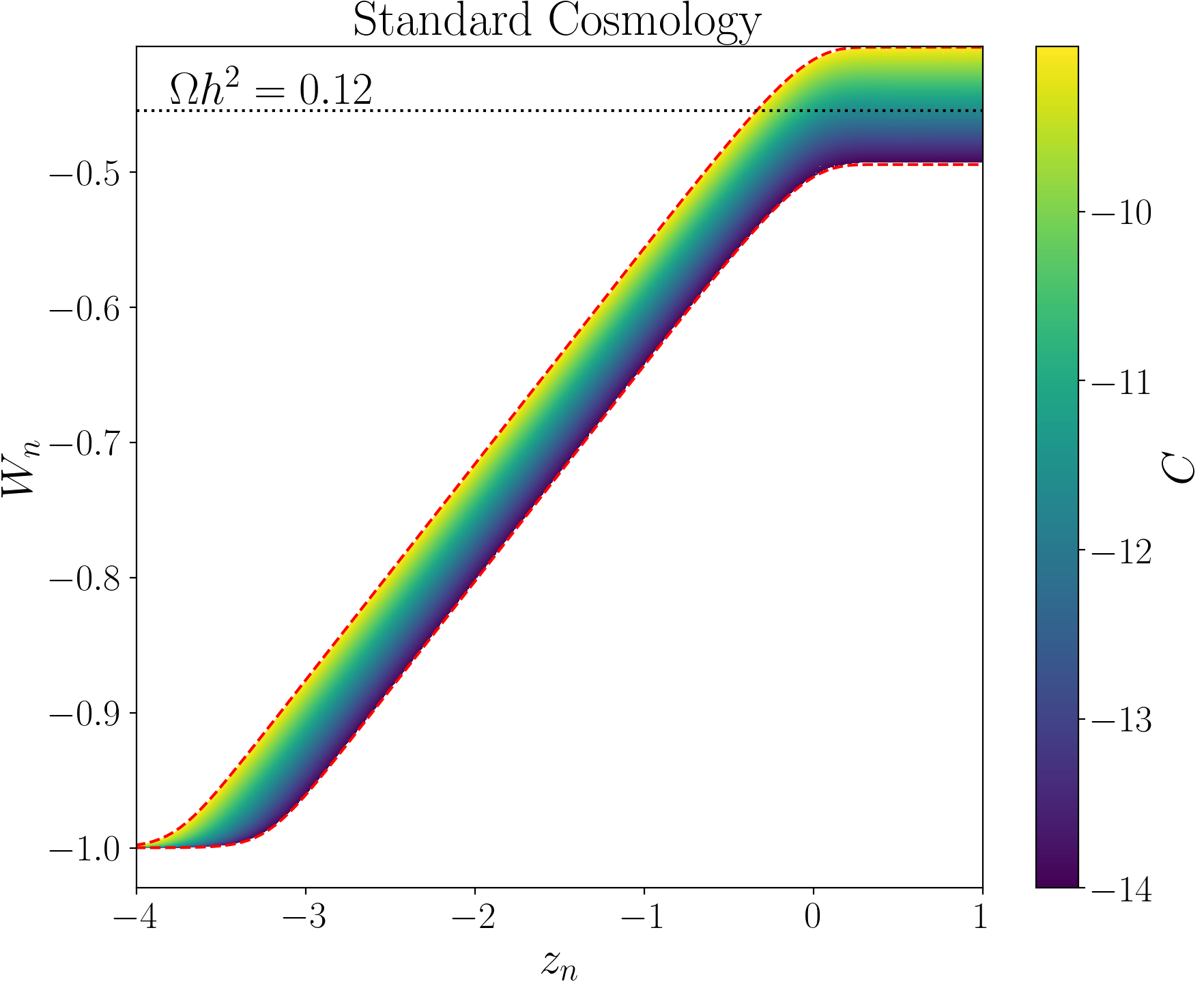}  \\ \includegraphics[scale=0.35,trim={0.0cm 0.0cm 3.5cm 0.0cm},clip]{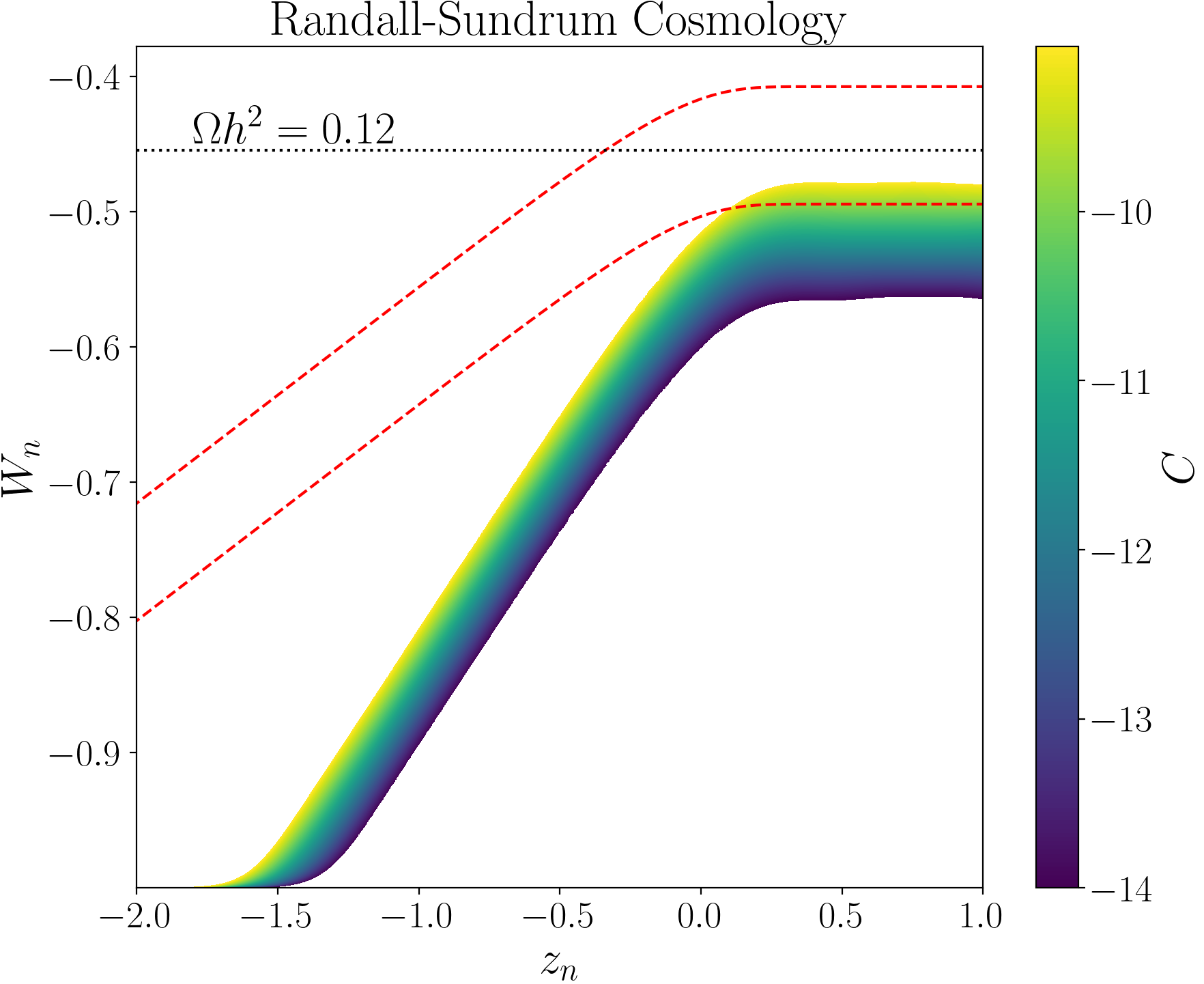} \hspace{+0.1cm} \includegraphics[scale=0.35,trim={0.0cm 0.0cm 3.5cm 0.0cm},clip]{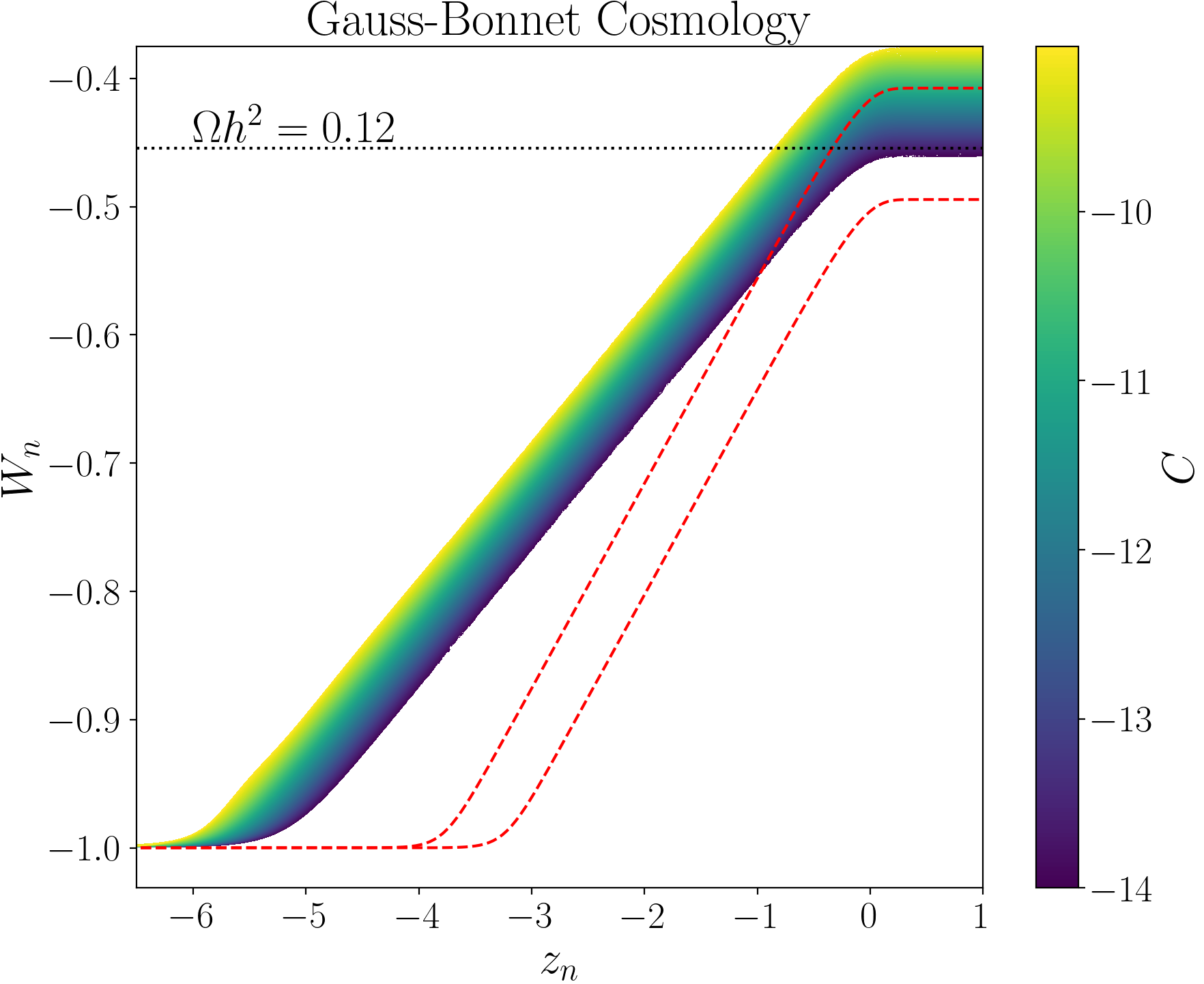}
        \caption{Parametric \textit{forward} PINN solution: particle yield evolution $W_n(z_n)$ for Standard (top), RS (bottom left) and GB (bottom right) cosmology. Parametric dependence of $W_n(z_n)$ on $C$ is shown via a colormap. The red-dashed contours represent FEM results for \( C = -14 \) and \( C = -9 \) in Standard cosmology across all panels. Horizontal dotted-black line corresponds to $W_n$ value satisfying $\Omega h^2 = 0.12$~\cite{Planck:2018vyg}.}
    \label{fig:parametricforward_yield}
    \end{figure*}
    \begin{figure*}[t!]
        \centering
        \includegraphics[scale=0.4]{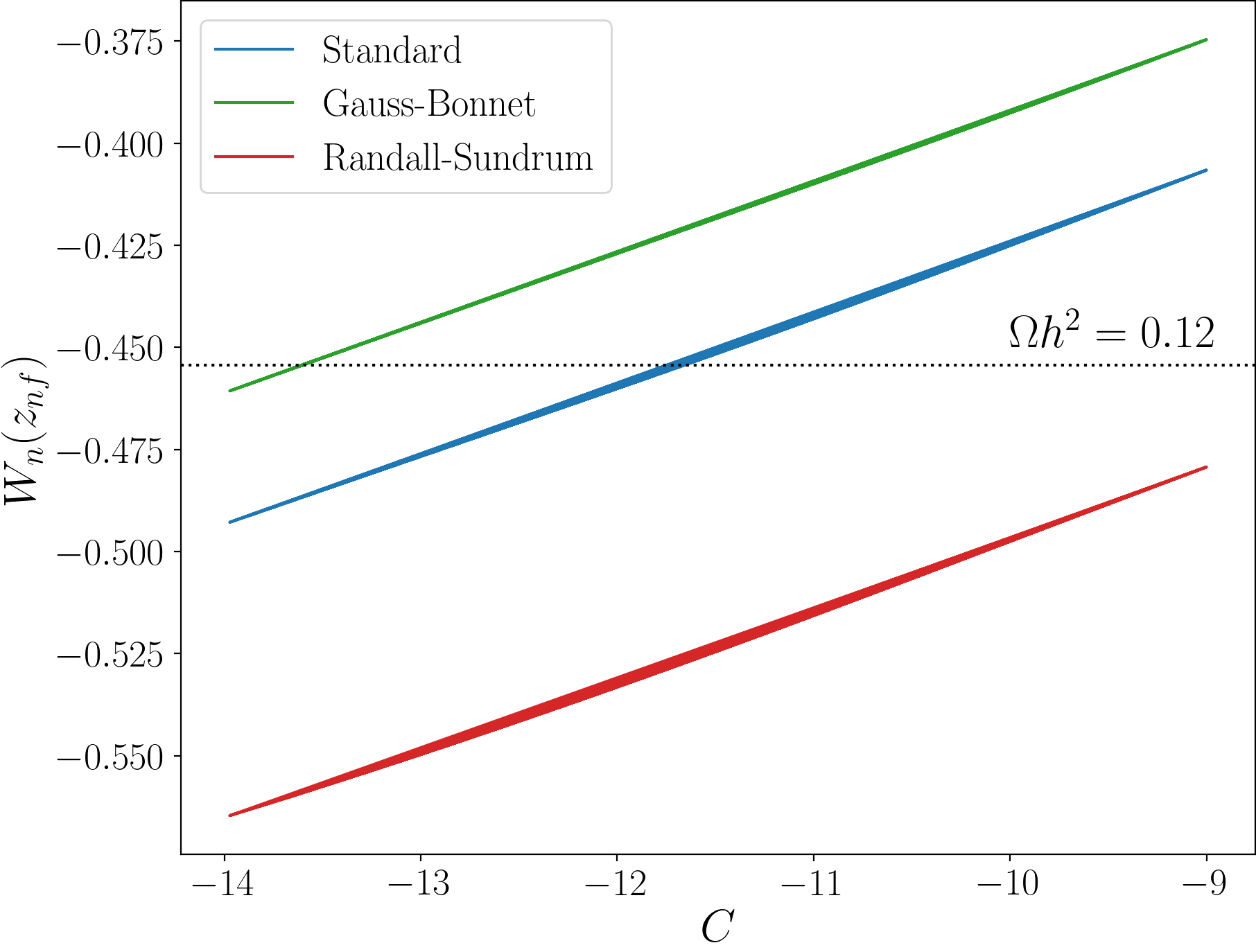}
        \caption{Relation between $W_n(z_{n f})$ and $C$ for Standard (blue), GB (green) and RS (red) cosmology.}
    \label{fig:parametricforward_linear}
    \end{figure*}
In Fig.~\ref{fig:parametricforward_yield} we present the parametric PINN solution for the \textit{forward} problem in Standard (top), RS (bottom left) and GB (bottom right) cosmology. The behavior of the yield $W_n(z_n)$ with respect to $C$ is represented by a colormap, while the red-dashed lines show the solution provided by \texttt{solve\textunderscore ivp}, when $C=-14$ and $C=-9$, for Standard cosmology. We observe a nearly perfect agreement between the PINN and \texttt{solve\textunderscore ivp} in the top plot, with the solution for $C \in [-14,-9]$ filling practically all the region between the red-dashed lines. This showcases the accuracy of the \textit{forward} PINN in solving the BE. Moreover, we notice that for $C \sim -11.5$, in Standard cosmology, our FIMP DM candidate accounts for the observed DM relic abundance (horizontal dotted-black line). For GB cosmology one would need a decreased value of $C \sim -13.5$ (bottom right plot), while for RS an increased value $C > -9$ is required (bottom left). These results are consistent with the theoretical expectations outlined in Section~\ref{sec:physics} [see Eqs.~\eqref{eq:Oh2Std} and~\eqref{eq:Oh2approxNS}]. In addition, in Fig.~\ref{fig:parametricforward_linear}, we present the yield value provided by the PINN for the last collocation point $W_n(z_{n f})$ in terms of $C$, for the three cosmologies. The PINN confirms the expected linear relation between relic density and cross-section for FIMP DM [see Eqs.~\eqref{eq:Oh2Std} and~\eqref{eq:Oh2approxNS}]. Overall, compared to FEMs, the \textit{forward} PINN provides a mesh-free method to solve the BE parametrically for cross-section values within an interval. Indeed, the NN is here a continuous function that approximates our target~$W(z, C)$.

    \begin{figure*}[t!]
        \centering
         \includegraphics[scale=0.3]{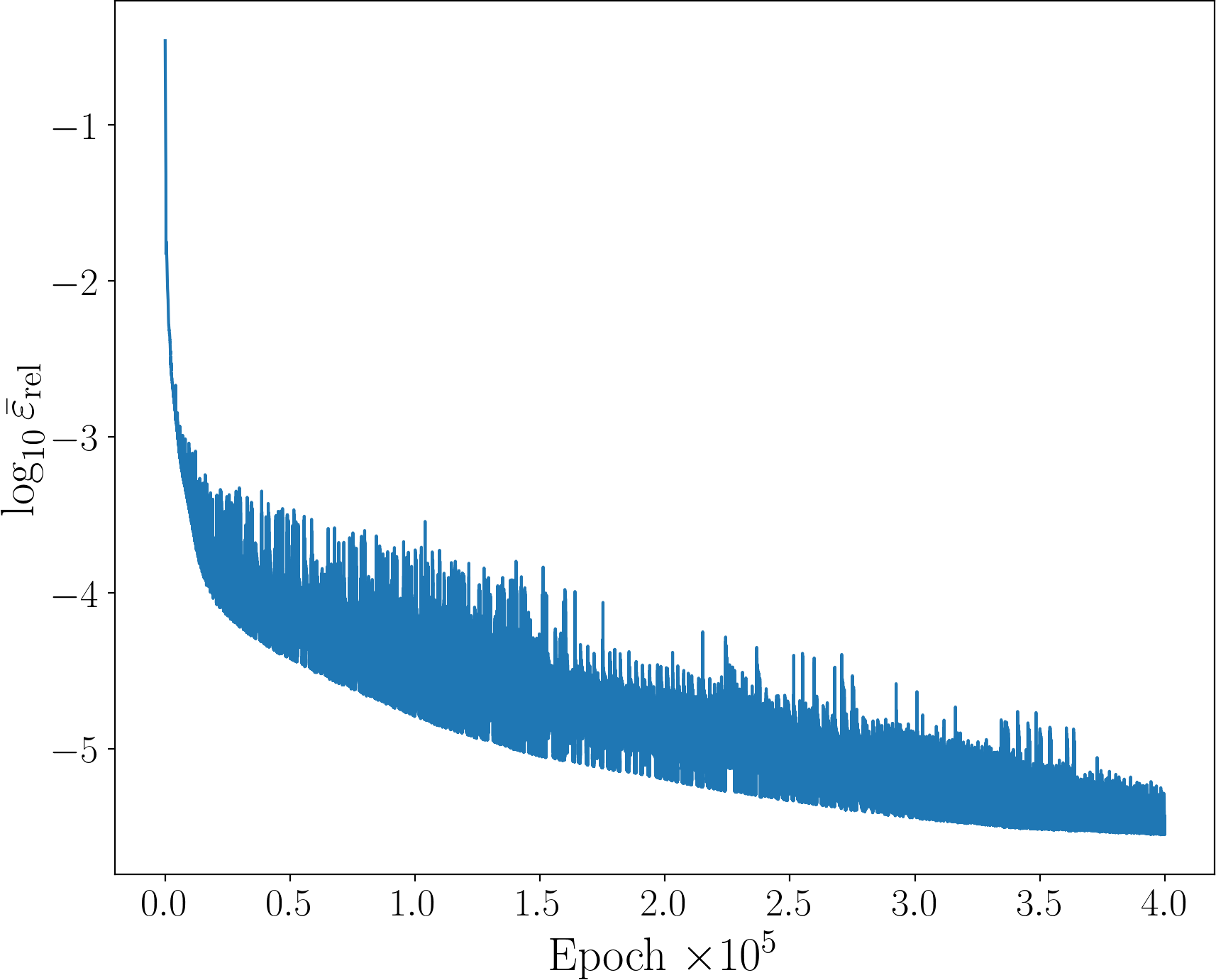} \hspace{+0.1cm} \includegraphics[scale=0.3]{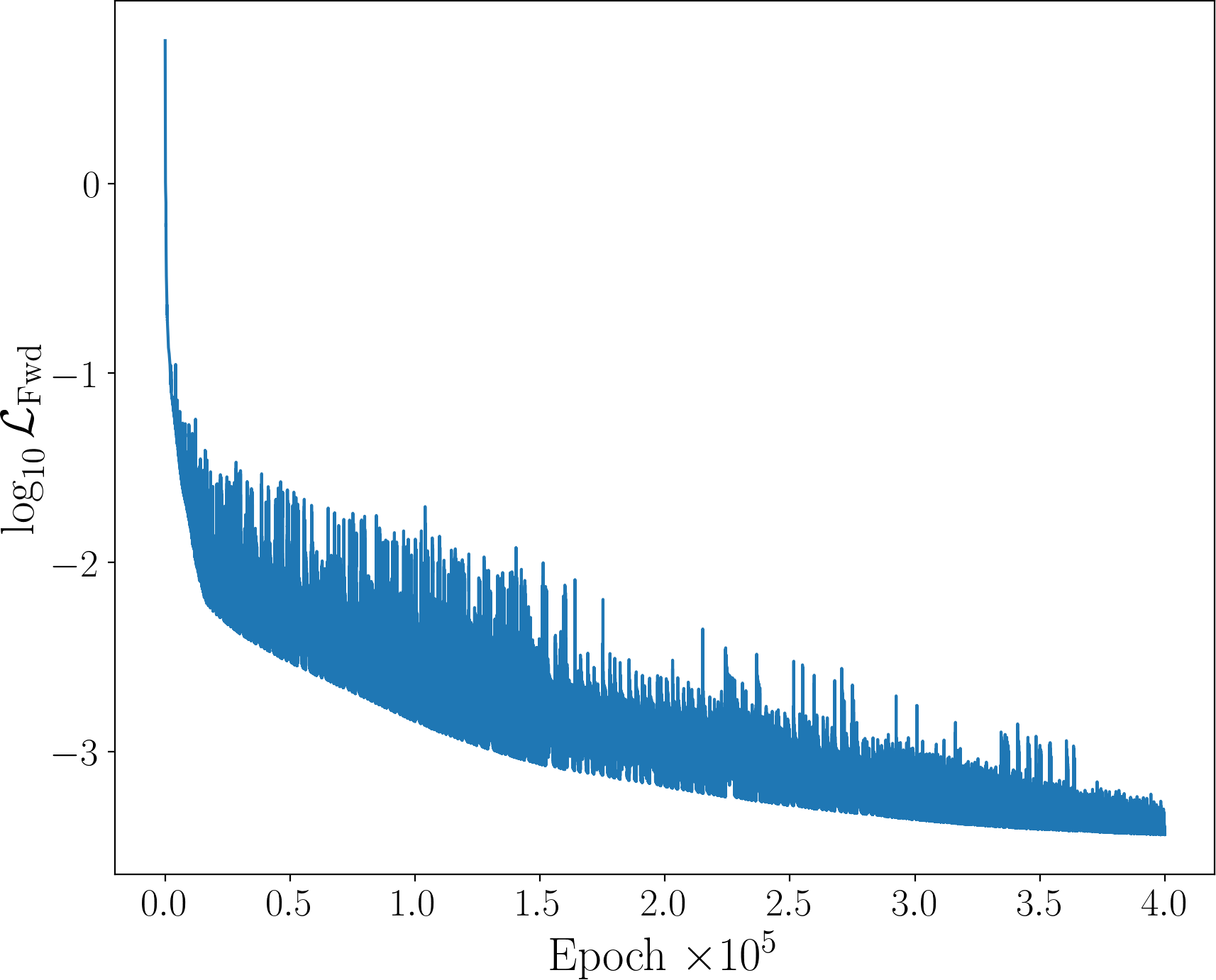}
        \caption{\textit{Forward} PINN solution: evolution in terms of epochs of the mean relative absolute error between PINN and FEM $W(z)$ results [see Eq.~\eqref{eq:MRAE}] (left) and total loss function $\mathcal{L}_{\text{Fwd}}$ [see Eqs.~\eqref{eq:Lfwd},~\eqref{eq:Lbe} and~\eqref{eq:Lic}] (right), for Standard cosmology with $C=-11.6964$ (see Sec.~\ref{sec:inverseC}).}
    \label{fig:FEMvsPINN}
    \end{figure*}
To systematically benchmark the performance of the \textit{forward} PINN against the FEM solver -- which we take as our ground truth -- it is useful to compute the mean relative absolute error:
\begin{equation}
    \bar{\varepsilon}_{\text{rel}} = \frac{1}{N_z} \sum_{j=1}^{N_z} \left| \frac{W(z_j)^{\text{PINN}} -W(z_j)^{\text{FEM}}}{W(z_j)^{\text{FEM}}} \right| \; ,
    \label{eq:MRAE}
\end{equation}
between the PINN and FEM solution for the particle yield $W(z)$. This allows to impose an accuracy threshold on the PINN results defining a systematic criterion to assess convergence and stop training. As an example, we present in Fig.~\ref{fig:FEMvsPINN} the \textit{forward} PINN solution for Standard cosmology with $C=-11.6964$ (see Sec.~\ref{sec:inverseC}). Namely, on the left we plot the evolution in terms of epochs of the mean relative absolute error defined above, where it is clear that at the end of training our PINN reaches sub-- $0.001\%$ agreement with the FEM solution. This result is consistent with the observed stabilization of the total PINN loss shown in the right plot, both plots exhibiting the same decreasing behavior in terms of the epochs. This is due to the fact that the mean relative absolute error is related to the residuals components of the loss function. These results showcase that in contrast to conventional NNs, where overfitting is a concern, PINNs typically improve with continued training. 

Although we focus here on a benchmark with constant $\langle \sigma v \rangle$ to solve the freeze-in BE using PINNs, it is useful to outline possible future extensions towards more realistic scenarios. Namely, a natural avenue for future research would be to incorporate full Boltzmann collision integrals, which are essential for modeling realistic interactions involving multi-particle states and complex processes such as coannihilation, semi-annihilation, and inelastic scatterings. These arise in a wide range of well-motivated concrete particle DM models, for e.g. involving the scalar portal~\cite{Yaguna:2011qn}, requiring dedicated numerical tools such as the state-of-the-art packages \texttt{MicrOmegas}~\cite{Belanger:2014vza,Belanger:2018ccd} and \texttt{DarkSUSY}~\cite{Bringmann:2018lay} to perform these computations. Thus, in order for PINNs to become a viable alternative to these widely used tools, future works should focus on incorporating the full Boltzmann collision integrals. As a first step with PINNs, it would be interesting to study one of the simplest realistic cases involving the thermally averaged cross section for $2 \to 2$ annihilation processes~\cite{Kolb:1988aj}:
\begin{equation}
\langle \sigma v \rangle(x) = \frac{g^2}{64 \pi^4} \left( \frac{m}{x} \right) \frac{1}{Y_{\mathrm{eq}}^2} \int_{4m^2}^{\infty} ds\, \sigma v(s)\, s\, \sqrt{s - 4m^2}\, \mathcal{K}_1\left( \frac{x \sqrt{s}}{m} \right) \; ,
\end{equation}
where $\sigma v(s)$ is the energy-dependent annihilation cross section which depends on the particle masses and interactions of a given model. One promising idea is to use a modular PINN approach~\cite{markidis2024braininspiredphysicsinformedneuralnetworks} where we construct a dedicated network to learn the thermal integral shown above while a second network solves the BE for particle DM abundance. Namely, as an input the first network would receive $\sigma v(s)$ and learn how to model $\langle \sigma v \rangle(x)$. The latter would then enter the freeze-in BE which is solved by the second PINN. This separation into multiple PINNs could facilitate more targeted training and efficient inference of $\langle \sigma v \rangle(x)$ and $Y(x)$, potentially offering a scalable and flexible framework for exploring more realistic DM scenarios with full collision dynamics.

\section{Solving \textit{inverse} problems with PINNs}
\label{sec:inverse}

%
    \begin{figure*}[t!]
        \centering
        \includegraphics[scale=0.095]{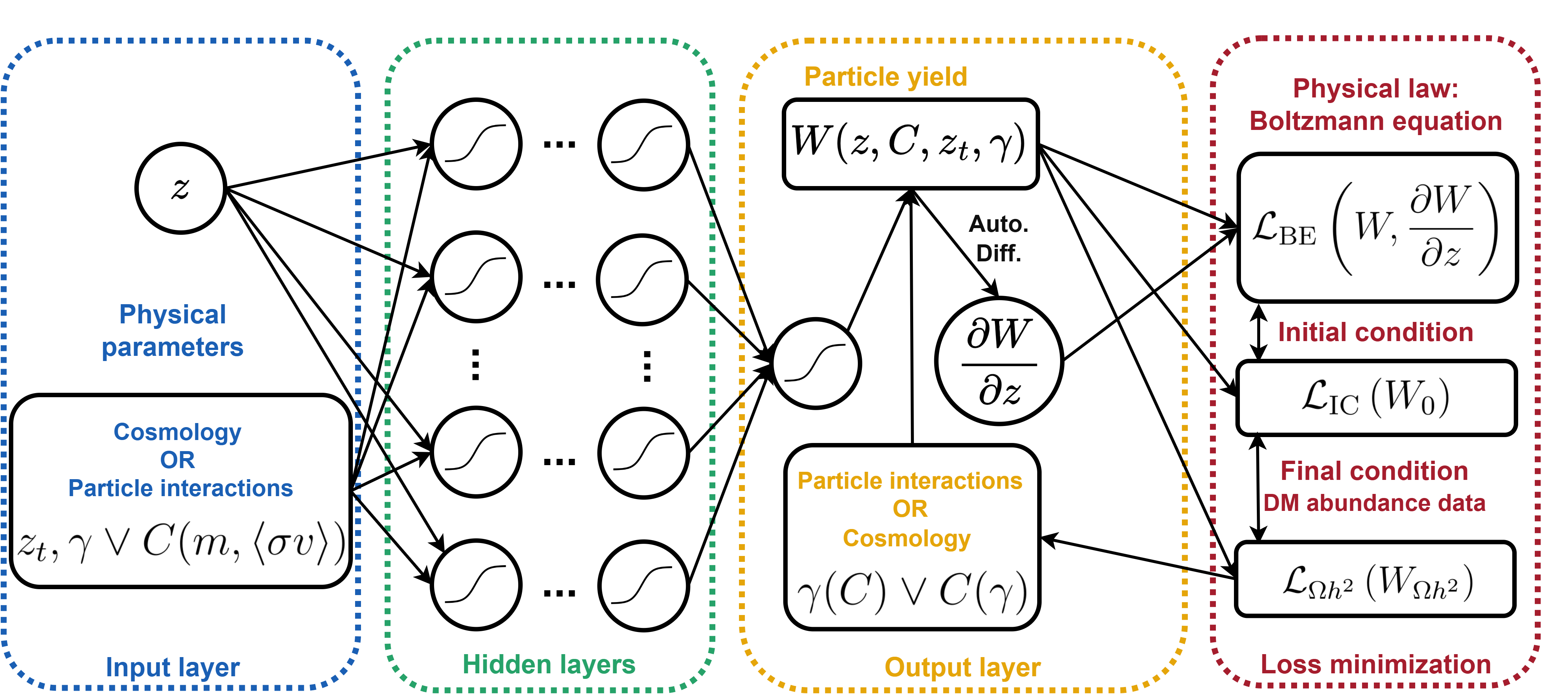}
        \caption{Schematic representation of the PINN structure for tackling \textit{inverse} problems in modeling the BE for freeze-in DM particle yield in alternative cosmology (see text for details).}
    \label{fig:inverse_PINN}
    \end{figure*}

PINNs are a particularly powerful technology in what regards \textit{inverse} problems. Our objective is to show that \textit{inverse} PINNs can be a valuable tool in theoretical physics to be used as a model and/or parameter space finder. In contrast to the \textit{forward} problem, where all the parameters of the BE are provided to the PINN before training, in \textit{inverse} problems we provide the data point of CDM abundance with some parameters of the theory left to be found. Using the PINN structure schematically depicted in Fig.~\ref{fig:inverse_PINN}, we will tackle two different problems of this kind:
\begin{itemize}

\item In Section~\ref{sec:inverseC}, we want to find the particle interaction strength, given the cosmological model, such that the observed CDM is satisfied. To be more specific, assuming $z_t=\ln 10$ and $\gamma=0$ (Standard cosmology), $\gamma=-2/3$ (GB cosmology) or $\gamma=2$ (RS cosmology), we want the PINN to find $C(m,\langle\sigma v\rangle)$ -- now promoted to a trainable parameter -- that solves Eq.~\eqref{eq:freezein2} and accounts for the observed DM relic density.

\item In Section~\ref{sec:inversegamma}, our goal is to find the power-law exponents, given the particle interaction strength, such that the observed CDM data point is obtained. Now it is $\gamma$ that is promoted to a trainable parameter whose value must be discovered by the PINN for multiple $C(m,\langle\sigma v\rangle)$ values. Considering again $z_t=\ln 10$, we want to find the pair-wise relation between $\gamma$ and $C(m,\langle\sigma v\rangle)$, that solves Eq.~\eqref{eq:freezein2} and explains the observed DM relic density value.

\end{itemize}
Furthermore, in Section~\ref{sec:inversetransition}, we will show how the \textit{inverse} PINN can adapt for a different parameterization of alternative cosmologies by replacing the switch-like transition function of Eq.~\eqref{eq:Fnew} with a smooth function. We will also use a Bayesian method in Section~\ref{sec:bayesian}, in order to quantify the epistemic uncertainty in $C$.

\subsection{Discovering particle interaction cross sections}
\label{sec:inverseC}
%
    \begin{figure*}[t!]
        \centering
        \includegraphics[scale=0.3]{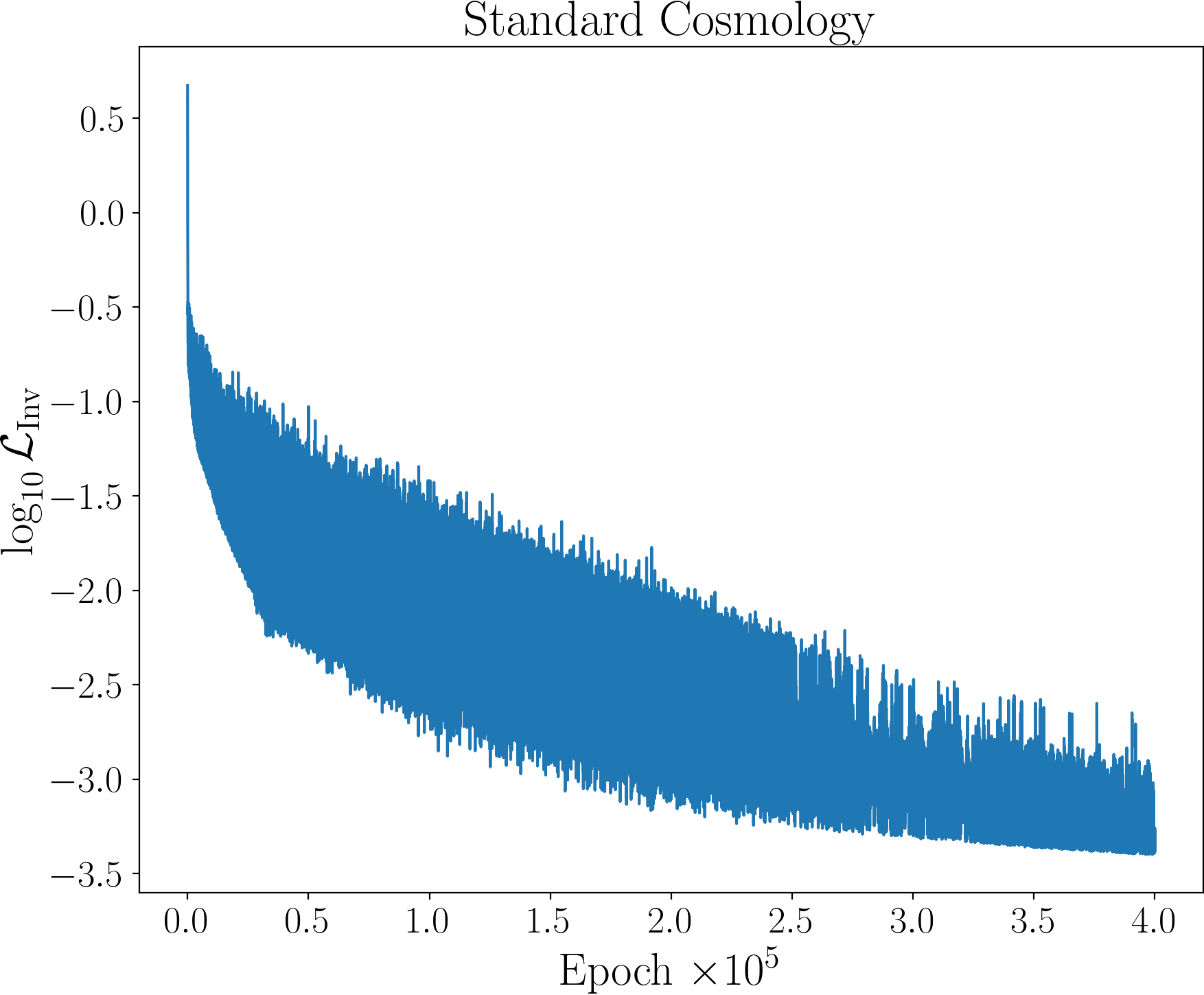}  \\ \includegraphics[scale=0.3]{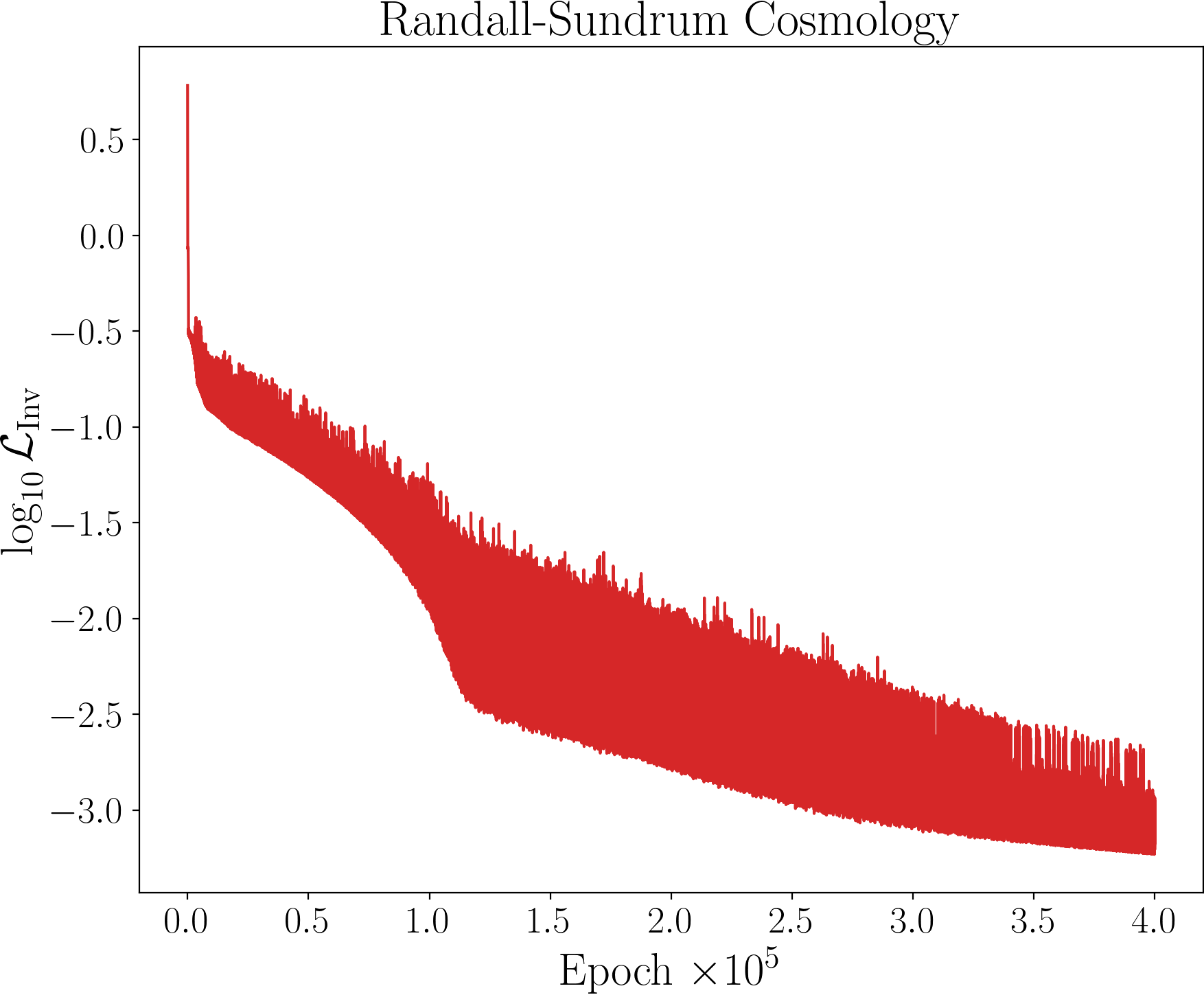} \hspace{+0.1cm} \includegraphics[scale=0.3]{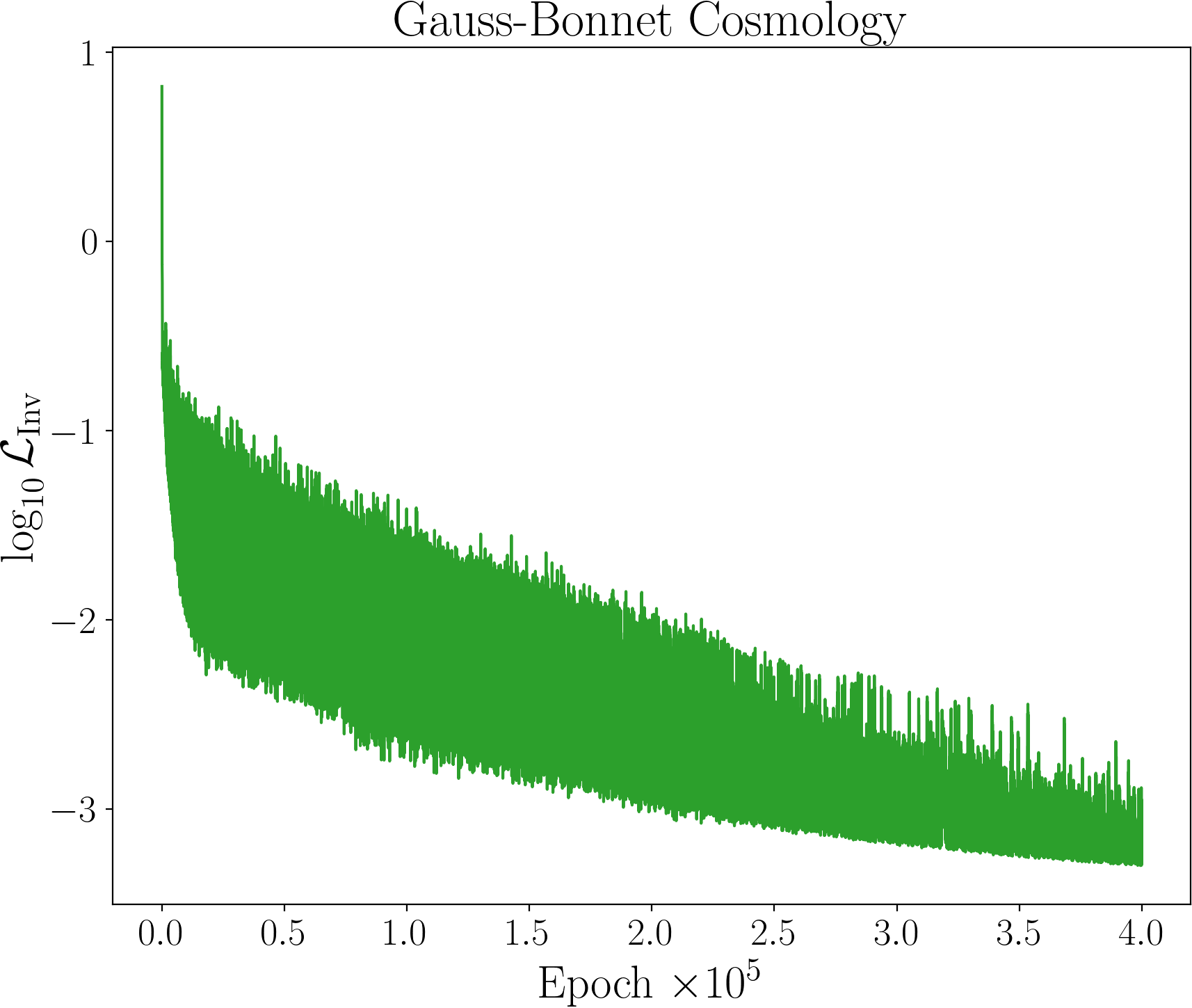}
        \caption{Evolution of the total \textit{inverse} problem loss function $\mathcal{L}_{\text{Inv}}$ [see Eqs.~\eqref{eq:Linv} and~\eqref{eq:LOh2}], for the cases of Standard (top), RS (bottom left) and GB (bottom right) cosmology.}
    \label{fig:inverse_losses}
    \end{figure*}

The information about the observed relic density enters \textit{inverse} PINNs in the form of a new term in the loss function. Thus, we minimize
\begin{equation}
    \mathcal{L}_{\text{Inv}} = \mathcal{L}_{\text{Fwd}} + \lambda_{\Omega h^2} \; \mathcal{L}_{\Omega h^2} \; ,
    \label{eq:Linv}
\end{equation}
where
\begin{equation}
    \mathcal{L}_{\Omega h^2} = \left|W(z_f) -W_{\Omega h^2}\right| \; ,
    \label{eq:LOh2}
\end{equation}
measures the absolute difference between the prediction of the PINN for the last collocation point $W(z_f)$ and the observed DM relic density [see Eq.~\eqref{eq:variablesnum}]. Like $\lambda_{\text{BE}}$ and $\lambda_{\text{IC}}$, the weight $\lambda_{\Omega h^2}$ is an hyperparameter that should be tuned to optimize the PINN's performance. As in the \textit{forward} problem we have previously discussed, no improvements were found by making those loss weights adaptive. Manual adjustment leads us to set $\lambda_\text{BE}=1$ and $\lambda_\text{IC}=\lambda_{\Omega h^2}=10$.

The collocation points in this problem are simply the values of $z$ in which we want to solve Eq.~\eqref{eq:freezein2}. We generate a total of 500 points, equally-spaced within $z \in [\ln 10^{-14},\ln 100]$. As for $C$, the variable that must be determined by the PINN, we initialize it following the normal prior
\begin{equation}
C\sim\mathcal{N}\left(-11.5,1.25\right)\;,
\label{eq:Cprior}
\end{equation}
meaning that its initial value stays within the interval $[-14,-9]$ with a probability close to $95\%$. Due to $C$ being trainable, there is a gradient being propagated through the PINN that determines how it changes at the end of each epoch. Consequently, we introduce a new learning rate, $\eta_C$, which acts exclusively on updating $C$. We found a good performance choosing $\eta_C = \eta$, both with a starting value of $10^{-3}$ and a decay rate of $0.99$ with a period of 1000 epochs. 

Fig.~\ref{fig:inverse_losses} displays the evolution of the loss function when the PINN was trained to find $C$ for Standard, RS and GB cosmologies. While it is clear that the loss is being minimized in all the three plots, it is important that we make some comments on the oscillations of $\mathcal{L}_{\text{Inv}}$, which are much larger than those of $\mathcal{L}_{\text{Fwd}}$ observed in Fig.~\ref{fig:parametricforward_losses}. The results obtained in Fig.~\ref{fig:parametricforward_linear} let us anticipate that the three cosmologies must have very different values of $C$ to accommodate the observed relic density. This raises some challenges on the adjustment of the learning rates. On the one hand, there are serious risks of making gradient descent diverge if those two hyperparameters are very large. On the other hand, the PINN has to be flexible enough to find $C$ independently of its initial value and cosmology, discouraging small learning rates. In comparison with the way we tuned the learning rate in Section~\ref{sec:forward}, note that its initial value is now 10 times larger. Regardless, there are always oscillations in the value of $C$ during training, which affect the value of the loss. These two factors explain, for the most part, the difference in the oscillations of the loss function in the \textit{inverse} and \textit{forward} problems. 

    \begin{figure*}[t!]
        \centering
        \includegraphics[scale=0.3]{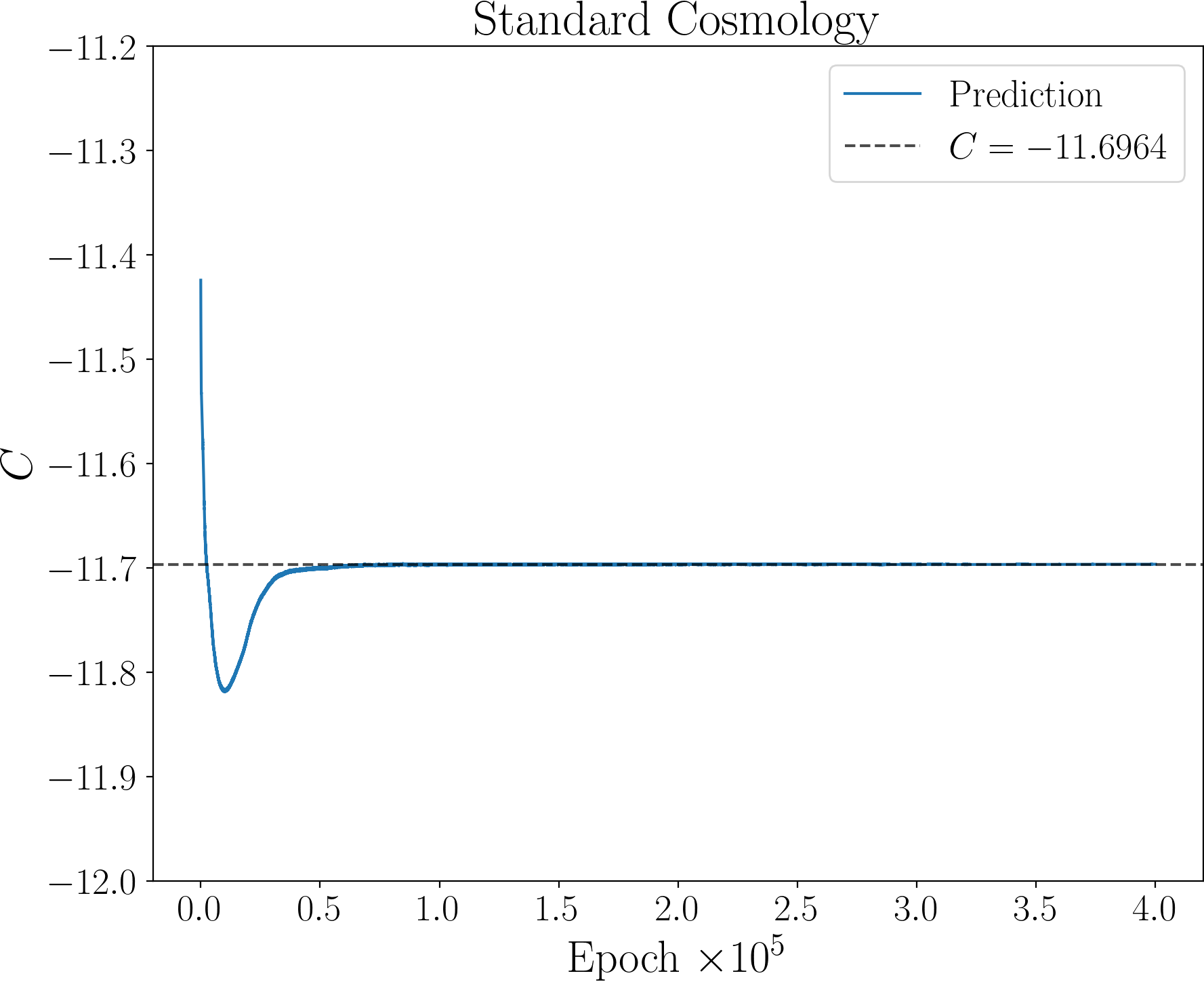}  \\ \includegraphics[scale=0.3]{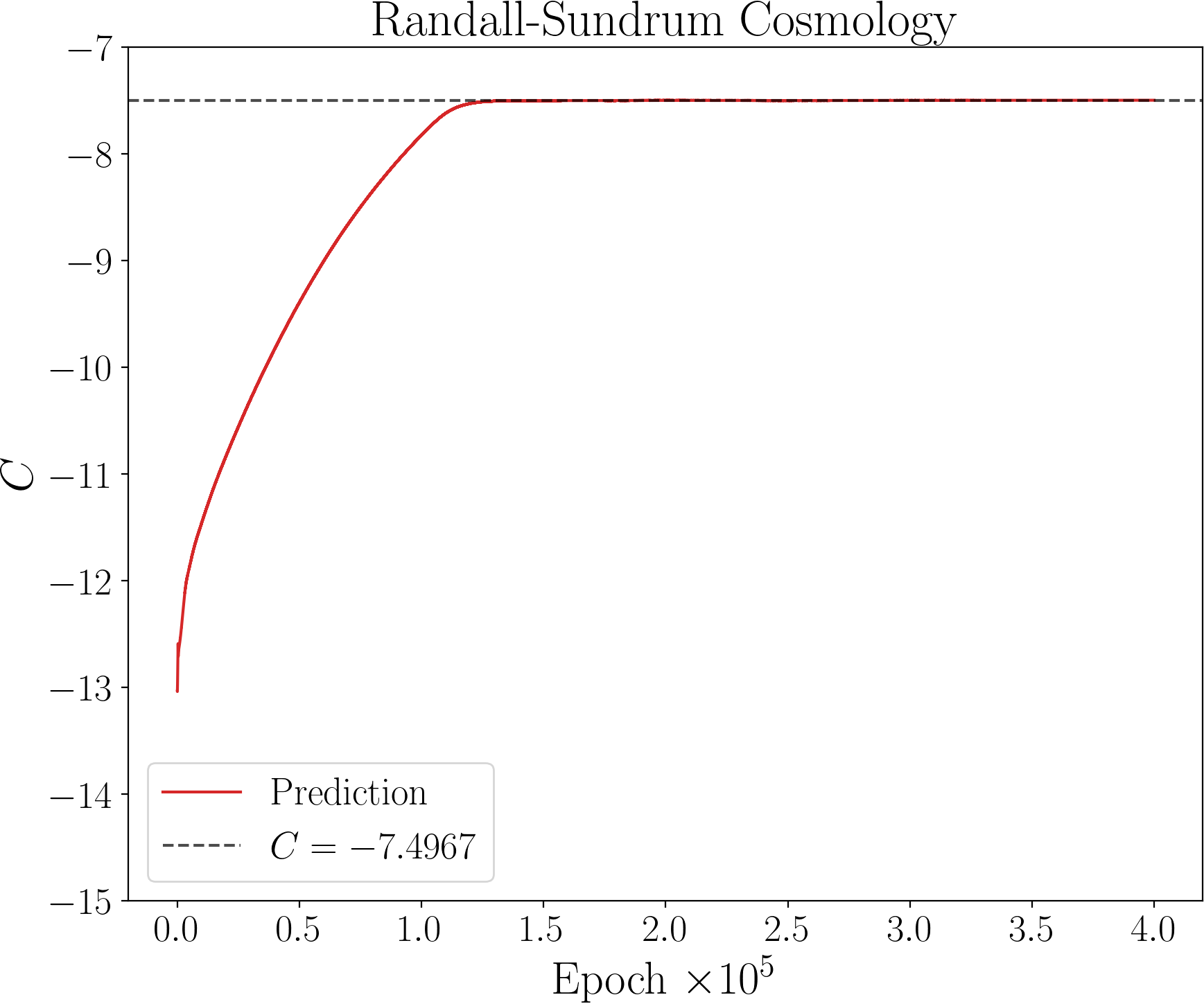} \hspace{+0.1cm} \includegraphics[scale=0.3]{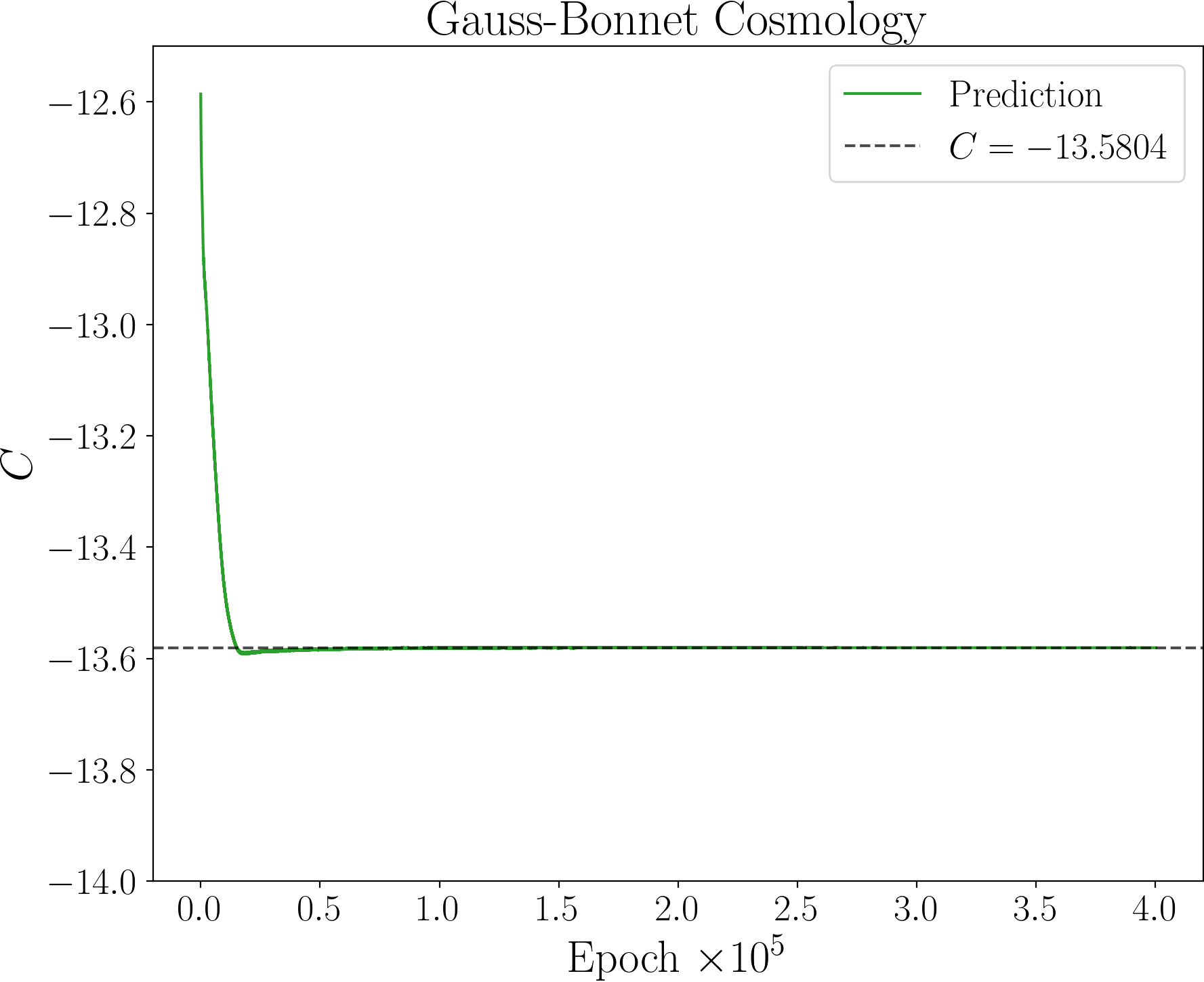}
        \caption{Evolution over epochs of the $C$ value determined by \textit{inverse} PINN, for Standard (top), RS (bottom left) and GB (bottom right) cosmology.}
    \label{fig:inverse_Closses}
    \end{figure*}
    \begin{figure*}[t!]
        \centering
        \includegraphics[scale=0.4]{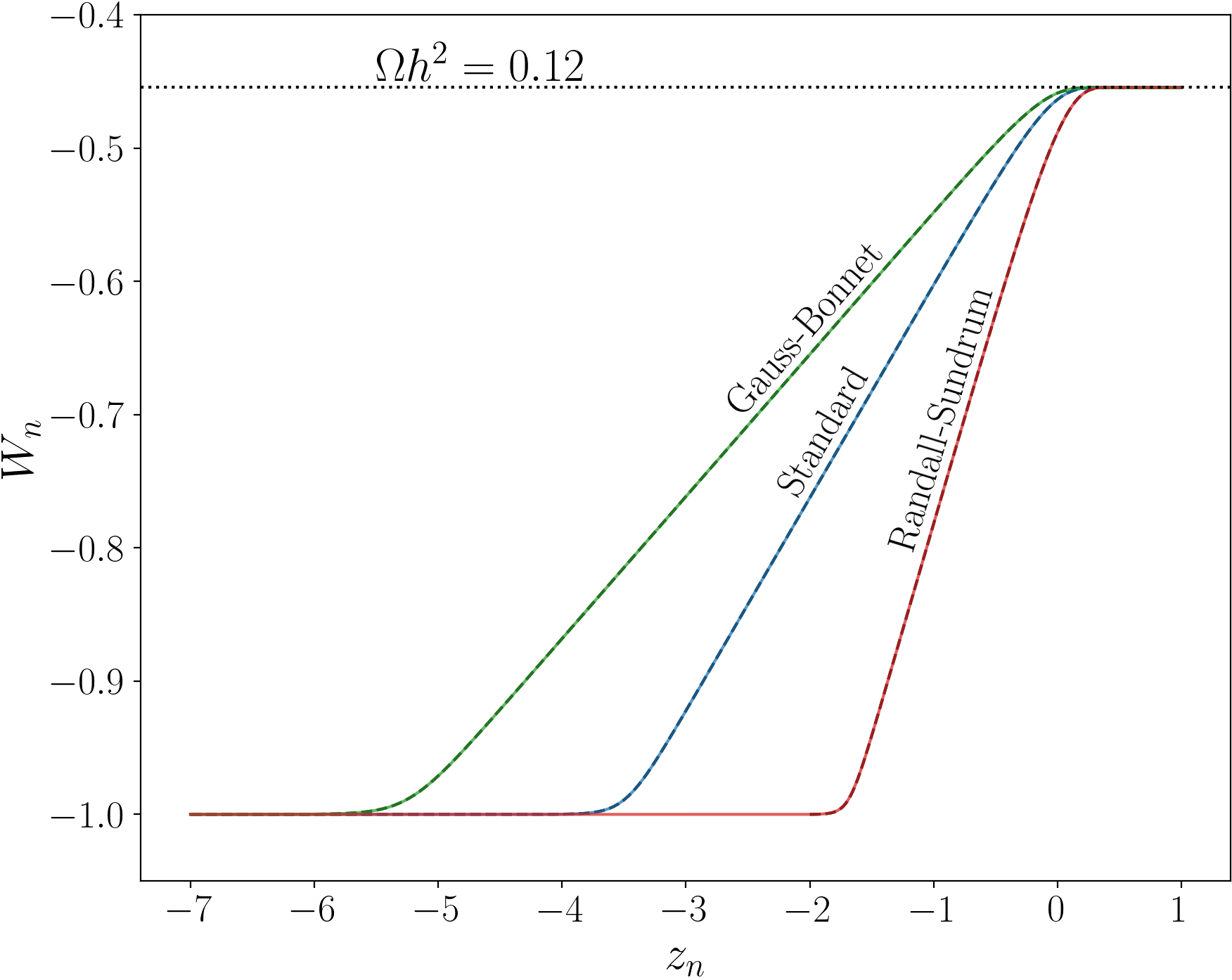}
        \caption{\textit{Inverse} PINN solution: particle yield evolution $W_n(z_n)$ for Standard (top), RS (bottom left) and GB (bottom right) cosmology. Black-dashed curves were obtained via FEM.}
    \label{fig:inverse_predictions}
    \end{figure*}
In Figs.~\ref{fig:inverse_Closses} and~\ref{fig:inverse_predictions}, we present the PINN solution for the \textit{inverse} problem of finding the FIMP DM particle yield evolution and determining the particle interaction strength for Standard (blue line), RS (red line) and GB (green line) cosmology. Namely, in Fig.~\ref{fig:inverse_Closses}, we show how $C$ evolves throughout training. The \textit{inverse} PINN reaches the correct value for particle cross-sections after about $50-100$ thousand epochs for the three cosmologies being tested. Remarkably, Fig.~\ref{fig:inverse_Closses} also shows that this can be achieved regardless of the initial $C$ being quite close (as in the case of the Standard cosmology) or far (as it happens in the alternative cosmologies) from the value that the PINN is supposed to discover. 

For \textit{inverse} PINNs, a direct FEM comparison is nontrivial, as FEMs are not inherently designed for \textit{inverse} inference, requiring embedding into a computationally intensive parameter scan within a broader optimization or search framework. A full comprehensive comparison between PINNs and FEMs lies beyond the scope of this work and warrants a dedicated study. Nonetheless we wish to outline that a possible approach could involve constructing a fine mesh grid in parameter space [e.g., for $(z,C)$ or $(z,\gamma)$] and solve the BE at all the points of the grid to identify the parameter values that reproduce the data. However, this would transform the FEM into a brute-force parameter-scan algorithm, which lacks scalability and efficiency. Instead, a feasible benchmarking strategy that can be adopted is to reinsert the values of $C$ (or $\gamma$) obtained by the PINN into the FEM solver to compute the yield function $W(z,C)$ [or $W(z,\gamma)$]. This enables the evaluation of the mean relative absolute error, as defined in Eq.~\eqref{eq:MRAE}, in terms of the training epochs. In addition, the FEM-computed yield at the final collocation point, $W(z_{ f})_{\text{FEM}}$, can be compared to the target value $W_{\Omega h^2}$ corresponding to the observed relic abundance, as a function of the training epochs. This strategy should provide a robust and reproducible quantitative metric for validating the \textit{inverse} PINN solution. Instead, here we adopt a simple strategy to cross-check the PINN results which we now discuss. The PINN predictions for $C$ and corresponding cross-section [see Eq.~\eqref{eq:Cdef}] are gathered in Table~\ref{tab:inverse_predictions_gammaVsC_1} for the three cosmologies. We also show the particle yield values for the last collocation point $W(z_f)$ and relic density predictions stemming from our FEM BE solver when we provide it the $C$ values obtained by the PINN and their respective $\gamma$'s. This serves as a cross-check of the PINN results in obtaining the correct CDM relic abundance. In the last column of the table, the central value of the data is reproduced up to the fourth decimal place, showcasing the accuracy of the \textit{inverse} PINN. Moreover, from Fig.~\ref{fig:inverse_predictions}, we also conclude that all cosmological scenarios reproduce the data $\Omega h^2 = 0.12$ (horizontal black-dashed line).
\begin{table}[!t]
\centering
\renewcommand*{\arraystretch}{1.3}
\begin{tabular}{|c||c|c||c|c|}  
\hline
$\gamma$ & $C_{\text{PINN}}$ & $\langle \sigma v \rangle$ $\left(\text{GeV}^{-2}\right)$ & $W(z_f)_{\text{FEM}}$ & $\Omega h^2_{\text{FEM}}$\\
\hline
$-2/3$ & $-13.5806$ & $3.9245 \times 10^{-28}$ & $-26.1588$ & $0.1200$\\
$0$ & $-11.6963$ & $2.5825 \times 10^{-27}$ & $-26.1585$ & $0.1200$\\
$2$ & $-7.4961$ & $1.7225 \times 10^{-25}$ & $-26.1580$ & $0.1201$\\
\hline
\end{tabular}
\caption{\textit{Inverse} PINN solution: finding $C$ [or equivalently $\langle \sigma v \rangle$ -- see Eq.~\eqref{eq:Cdef}] for GB ($\gamma = -2/3$), Standard ($\gamma = 0$) and RS ($\gamma = 2$) cosmology -- see Figs.~\ref{fig:inverse_Closses} and~\ref{fig:inverse_predictions}. Last two-columns indicate relic abundance values obtained via FEM.}
\label{tab:inverse_predictions_gammaVsC_1}
\end{table}

In contrast to PINNs, FEMs are not inherently designed for solving \textit{inverse} problems and typically require external optimization methods. Furthermore, we note that the FEM solver is very sensitive to the IC values of $W_0$ evaluated at $z_0$, sometimes becoming unstable. This can be seen in Fig.~\ref{fig:inverse_predictions}, when we compare the starting point of the FEM (black-dashed) solution with the one of PINN (red) for RS. This problem does not occur with the PINN, showcasing them as a good alternative to solve IC and/or boundary and/or final condition problems in these type of ODEs.

\subsection{Discovering power-law cosmologies}
\label{sec:inversegamma}

%
    \begin{figure*}[t!]
        \centering
        \includegraphics[scale=0.4]{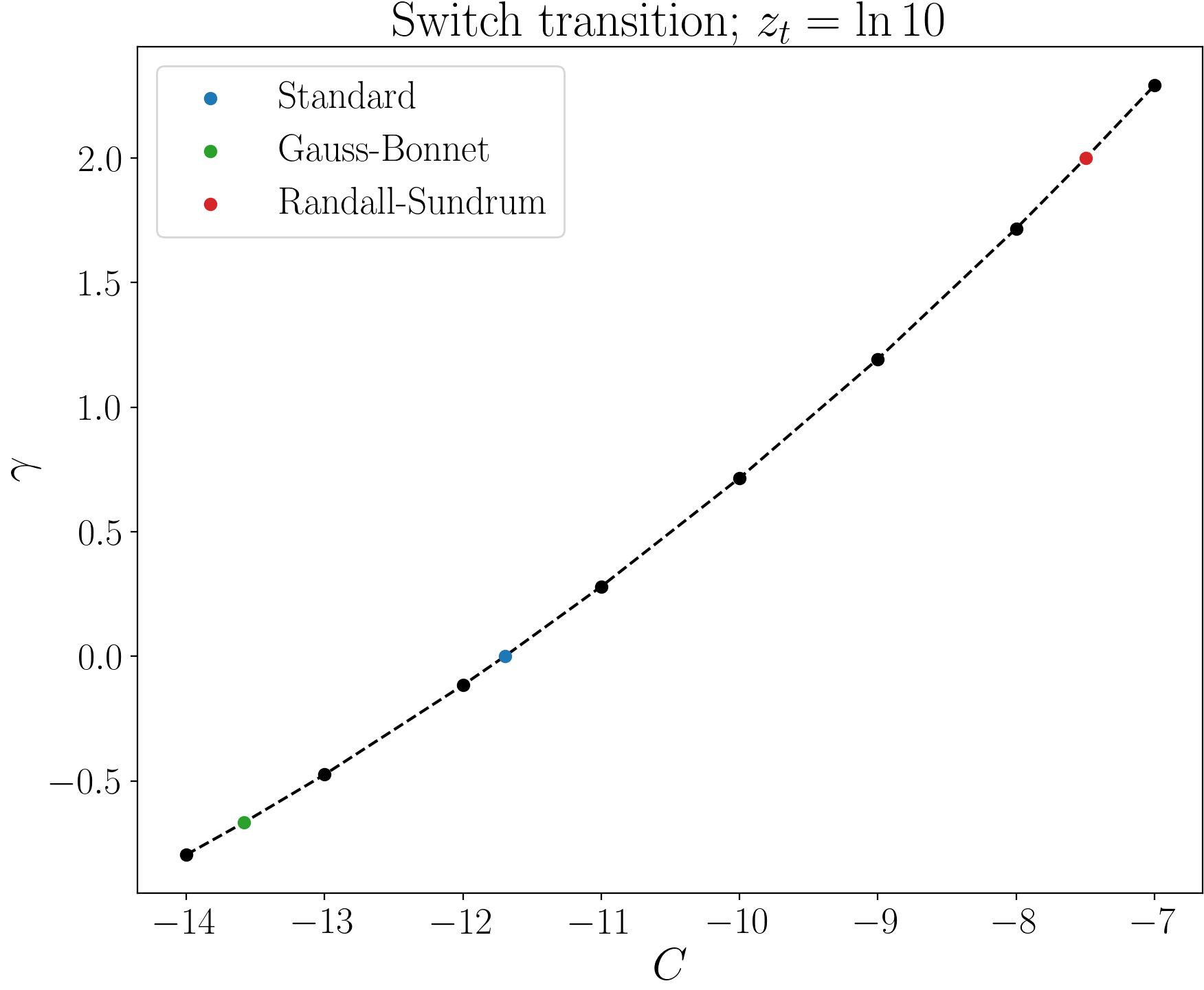}
        \caption{PINN solution for the \textit{inverse} problem (see text for details): relation between $\gamma$ and $C$, for $z_t = \ln 10$ and switch transition~\eqref{eq:Fnew} in BE~\eqref{eq:freezein2}, satisfying $\Omega h^2 = 0.12$~\cite{Planck:2018vyg}. The dashed-line was obtained via linear interpolation of the points. Colored point in green (blue) [red] indicate the case of GB (Standard) [RS] cosmology (see Figs.~\ref{fig:inverse_Closses} and~\ref{fig:inverse_predictions}), while the black point values are in Table~\ref{tab:inverse_predictions_gammaVsC_2}.}
    \label{fig:inverse_predictions_gammaVsC}
    \end{figure*}
\begin{table}[!t]
\centering
\renewcommand*{\arraystretch}{1.3}
\begin{tabular}{|c|c||c||c|c|}  
\hline
 $C$ & $\langle \sigma v \rangle$ $\left(\text{GeV}^{-2}\right)$ & $\gamma_{\text{PINN}}$ & $W(z_f)_{\text{FEM}}$ & $\Omega h^2_{\text{FEM}}$ \\
\hline
$-14$ & $2.5796 \times 10^{-28}$ & $-0.7969$ & $-26.1611$ & $0.1197$\\
$\mathbf{-13.5804}$ & $3.9245 \times 10^{-28}$ & $\mathbf{-0.6667}$ & $-26.1585$ & $0.1200$\\
$-13$ & $7.0120 \times 10^{-28}$ & $-0.4750$ & $-26.1588$ & $0.1200$\\
$-12$ & $1.9061 \times 10^{-27}$ & $-0.1162$ & $-26.1587$ & $0.1200$\\
$\mathbf{-11.6964}$ & $5.1812 \times 10^{-27}$ & $\mathbf{0.0001}$ & $-26.1588$ & $0.1200$\\
$-11$ & $2.5822 \times 10^{-27}$ & $0.2800$ & $-26.1587$ & $0.1200$\\
$-10$ & $1.4084 \times 10^{-26}$  & $0.7156$ & $-26.1586$ & $0.1200$\\
$-9$ & $3.8284 \times 10^{-26}$ & $1.1936$ & $-26.1588$ & $0.1200$\\
$-8$ & $1.0407 \times 10^{-25}$  & $1.7174$ & $-26.1586$ & $0.1200$\\
$\mathbf{-7.4967}$ & $1.7215 \times 10^{-25}$ & $\mathbf{2.0005}$ & $-26.1586$ & $0.1200$\\
$-7$ & $2.8289 \times 10^{-25}$ & $2.2922$ & $-26.1584$ & $0.1200$\\
\hline
\end{tabular}
\caption{PINN solution for the \textit{inverse} problem of finding  $\gamma$ given $C$, for $z_t = \ln 10$ and switch transition~\eqref{eq:Fnew} -- see Fig.~\ref{fig:inverse_predictions_gammaVsC}. Last couple of columns serve as a cross-check of PINN results with FEM.}
\label{tab:inverse_predictions_gammaVsC_2}
\end{table}
As we now address the problem of determining $\gamma$, we provide as input to the PINN a discrete set of~$C$ values $[-14,-13, \cdots, -8,-7]$, along with the ones that characterize the Standard, RS and GB cosmologies. The choice of initial value for $\gamma$ is based on the normal prior
\begin{equation}
    \gamma\sim\mathcal{N}\left(\frac{2}{3}, \frac{2}{3}\right) \;,
    \label{eq:gamma_prior}
\end{equation}
allowing it to be, with $95\%$ probability, within the interval bounded by the power-law parameters of the GB and RS cosmologies, namely $[-2/3, 2]$. No changes were made to the loss weights nor to the learning rates that we applied in the previous \textit{inverse} problem. We conclude from Fig.~\ref{fig:inverse_predictions_gammaVsC} that the PINN found a pair-wise relationship between the power-law exponent and interaction cross-sections, such that the observed CDM data is reproduced. This is further confirmed by the results organized in Table~\ref{tab:inverse_predictions_gammaVsC_2}. Once again, we cross-check the PINN's accuracy with a FEM. 

Before we move on to the next section, we should highlight that all results have been obtained for $z_t = \ln 10$ and switch transition parameterization~\eqref{eq:Fnew}. For this transition temperature regime $x_t = 10 \gg 1$, the analytical approximation of Eq.~\eqref{eq:Oh2approxNS} holds. In fact, we checked that this approximation is in agreement with the exact PINN results beyond the fourth decimal digit in the relic abundance prediction, clearly for experimental purposes well within $\Omega h^2_{\text{CDM}} = 0.120 \pm 0.001$~\cite{Planck:2018vyg}. In what follows, we will take a closer look at the transition function modeling our alternative power-law cosmological scenarios. Specifically, we will test the PINN in a $z_t$ regime where this approximation is no longer valid and propose a new modified transition function.

\subsection{Modeling alternative cosmology: transition between power and Hubble law}
\label{sec:inversetransition} 

%
    \begin{figure*}[t!]
        \centering
        \includegraphics[scale=0.4]{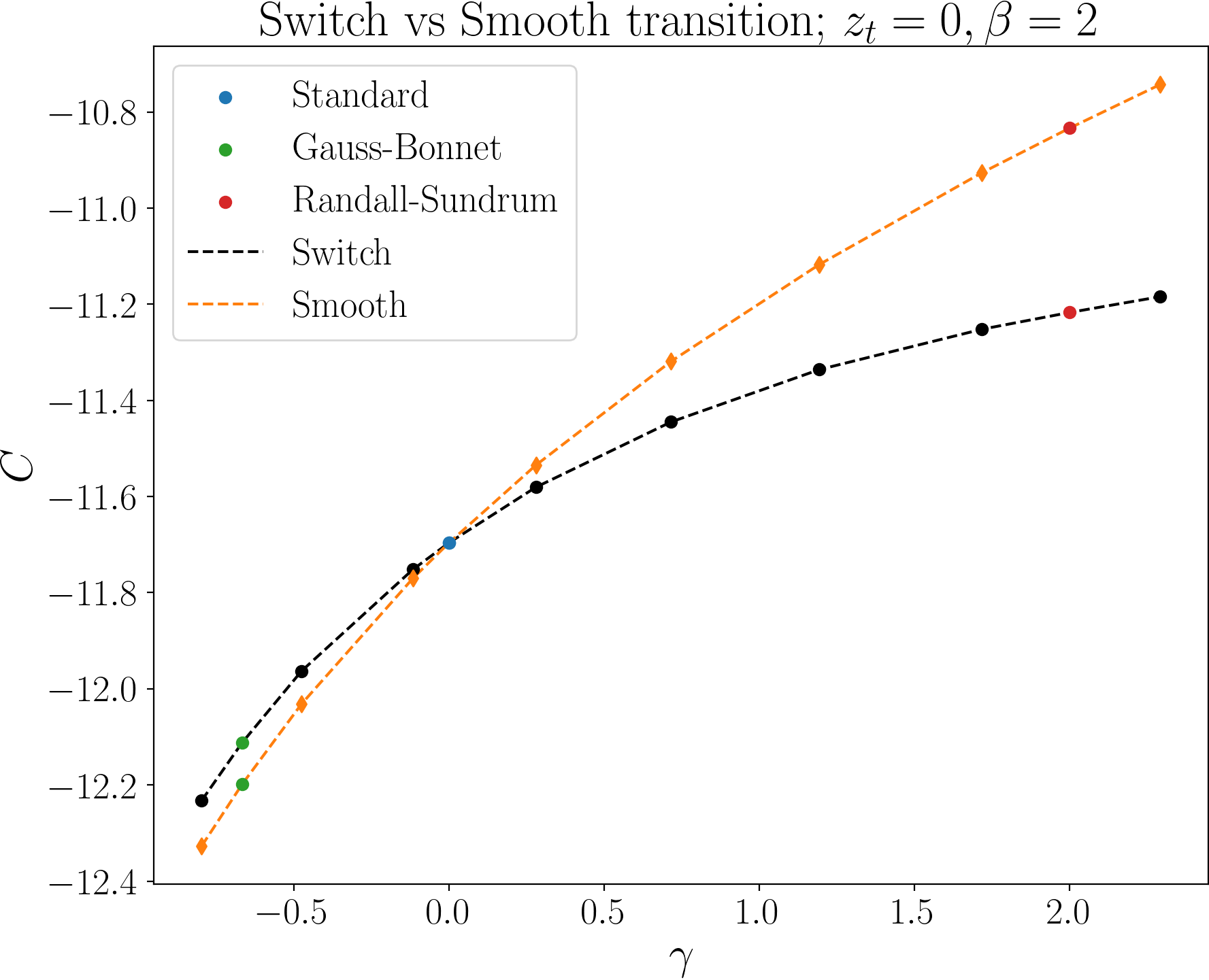}
        \caption{\textit{Inverse} PINN solution: relation between $\gamma$ and $C$, for $z_t = 0$ and switch transition of BE~\eqref{eq:freezein2} (black-dashed curve), as well as smooth transition of BE~\eqref{eq:freezein_smooth} with $\beta = 2$ (orange-dashed curve), satisfying $\Omega h^2 = 0.12$~\cite{Planck:2018vyg}. Dashed-lines were obtained via linear interpolation of the points. Colored point in green (blue) [red] indicate the case of GB (Standard) [RS] cosmology, while the black and orange point values are gathered in Tables~\ref{tab:inverse_predictions_gammaVsC_switch} and~\ref{tab:inverse_predictions_gammaVsC_smooth}, respectively.}
    \label{fig:inverse_predictions_swithcVSsmooth}
    \end{figure*}
Up to now, we have discussed alternative cosmological scenarios through the parameterization function $F(x,x_t,\gamma)$,
defined in Eq.~\eqref{eq:Hpowerlaw}. The transition between power and Hubble law expansion of the Universe was modeled indeed as a switch, which enabled us to describe it through Eq.~\eqref{eq:Fnew} as a ReLU function. It is, however, interesting to consider a transition regime that is smooth, adding both complexity and realism to the cosmological model.

In this section, inspired by ReLU as a switch function, we consider the possibility of a modified
softplus function defined by
\begin{equation}
    \mathrm{Softplus}_\beta (x) := \frac{1}{\beta} \ln \left( 1 + e^{\beta x} \right) \; ,
\end{equation}
which exhibits the following useful properties:
\begin{align}
    &\lim_{\beta \rightarrow \infty} \mathrm{Softplus}_\beta (x) = \mathrm{ReLU}(x) = \max \left( 0, x \right) \; ,  \\[2mm]
    &\mathrm{Softplus}_\beta (x) \sim x \quad \text{as} \quad x \rightarrow \infty \, , \quad \forall \beta > 0 \; ,  \\[2mm]
    &\mathrm{Softplus}_\beta (x) \sim 0 \quad \text{as} \quad x \rightarrow - \infty \, , \quad \forall \beta > 0 \; ,  \\[2mm]
    &\left. \frac{\partial \mathrm{Softplus}_\beta (x)}{\partial x} \right|_{x=0} = \frac{1}{2}, \quad \forall \beta \; .
\end{align}
Hence, $\mathrm{Softplus}_\beta (x)$ is a smooth approximation to ReLU, where the ``speed" of the transition is controlled
by $\beta$, recovering the switch transition in the limit $\beta \rightarrow \infty$.
The middle point of this transition is fixed at $x=0$, and is independent of the choice of $\beta$.
Thus, we can define a new parameterization function -- $G(x, x_t, \gamma)$ -- that models the transition between power and Hubble law in alternative cosmology:
\begin{equation}
    G(x, x_t, \gamma; \beta) = 
    \exp \left[ \gamma \ \mathrm{Softplus}_\beta(z_t - z) \right] =
    \left[ 1 + \left( \frac{x_t}{x} \right)^\beta \right]^\frac{\gamma}{\beta} \; ,
    \label{eq:smooth_func}
\end{equation}
with
\begin{equation}
    \lim_{\beta \rightarrow \infty} G(x, x_t, \gamma; \beta) = F(x, x_t, \gamma) \; .
\end{equation}

The BE~\eqref{eq:freezein2} is now modified as follows,
\begin{equation}
    \mathcal{E}\left[z, W, \frac{d W}{dz} ; C, z_t, \gamma ; \beta \right] \equiv \frac{d W}{dz} - 
    \exp(C - \gamma \ \mathrm{Softplus}_\beta(z_t - z) - z + 2 W_{\text{eq}} - W) = 0 \; .
\label{eq:freezein_smooth}
\end{equation}
Here we make use of the loss function for \textit{inverse} PINN in Eq.~\eqref{eq:Linv} [see also Eq.~\eqref{eq:Lfwd}], featuring the above BE.

\begin{table}[!t]
\centering
\renewcommand*{\arraystretch}{1.3}
\begin{tabular}{|c||c|c||c|c|}  
\hline
 $\gamma$ & $C_{\text{Switch}}$ & $\langle \sigma v \rangle$ $\left(\times 10^{-27} \ \text{GeV}^{-2}\right)$ & $W(z_f)_{\text{FEM}}$ & $\Omega h^2_{\text{FEM}}$  \\
\hline
$-0.7969$ & $-12.2327$       & $1.5104$ & $-26.1596$ &  $0.1199$\\
$\mathbf{-2/3}$ & $-12.1125$ & $1.7033$ & $-26.1581$ &  $0.1200$\\
$-0.4750$ & $-11.9632$       & $1.9775$ & $-26.1567$ &  $0.1202$\\
$-0.1162$ & $-11.7519$       & $2.4428$ & $-26.1585$ &  $0.1200$\\
$\mathbf{0}$ & $-11.6964$    & $2.5822$ & $-26.1586$ &  $0.1200$\\
$0.2800$ & $-11.5808$        & $2.8986$ & $-26.1566$ &  $0.1202$\\
$0.7156$ & $-11.4450$        & $3.3202$ & $-26.1551$ &  $0.1204$\\
$1.1936$ & $-11.3360$        & $3.7026$ & $-26.1531$ &  $0.1207$\\
$1.7174$ & $-11.2521$        & $4.0267$ & $-26.1550$ &  $0.1204$\\
$\mathbf{2}$ & $-11.2172$    & $4.1697$ & $-26.1569$ &  $0.1202$\\
$2.2922$ & $-11.1843$        & $4.3092$ & $-26.1567$ &  $0.1202$\\
\hline
\end{tabular}
\caption{\textit{Inverse} PINN solution: determining $C$ given $\gamma$, for $z_t = 0$ and switch transition~\eqref{eq:Fnew} -- see Fig.~\ref{fig:inverse_predictions_swithcVSsmooth}. Values obtained through FEM serve as a benchmark of comparison with PINN results.}
\label{tab:inverse_predictions_gammaVsC_switch}
\end{table}
\begin{table}[!t]
\centering
\renewcommand*{\arraystretch}{1.3}
\begin{tabular}{|c||c|c||c|c|}  
\hline
 $\gamma$ & $C_{\text{Smooth}}$ & $\langle \sigma v \rangle$ $\left(\times 10^{-27} \ \text{GeV}^{-2}\right)$ & $W(z_f)_{\text{FEM}}$ & $\Omega h^2_{\text{FEM}}$ \\
\hline
$-0.7969$ & $-12.3274$     & $1.3739$ & $-26.1590$ & $0.1200$ \\
$\mathbf{-2/3}$ & $-12.1990$     & $1.5621$ & $-26.1588$ & $0.1200$\\
$-0.4750$ & $-12.0317$     & $1.8466$ & $-26.1585$ & $0.1200$\\
$-0.1162$ & $-11.7704$     & $2.3980$ & $-26.1587$ & $0.1200$\\
$\mathbf{0}$  & $-11.6965$     & $2.5819$ & $-26.1587$ & $0.1200$\\
$0.2800$  & $-11.5350$     & $3.0345$ & $-26.1588$ & $0.1200$ \\
$0.7156$  & $-11.3190$     & $3.7661$ & $-26.1588$ & $0.1200$ \\
$1.1936$  & $-11.1176$     & $4.6064$ & $-26.1587$ & $0.1200$\\
$1.7174$  & $-10.9265$     & $5.5764$ & $-26.1587$ & $0.1200$ \\
$\mathbf{2}$  & $-10.8337$     & $6.1187$ & $-26.1590$ & $0.1200$ \\
$2.2922$  & $-10.7429$     & $6.7002$ & $-26.1586$ & $0.1200$ \\
\hline
\end{tabular}
\caption{\textit{Inverse} PINN solution: determining $C$ given $\gamma$, for $z_t = 0$ and smooth transition~\eqref{eq:smooth_func} with $\beta = 2$ -- see Fig.~\ref{fig:inverse_predictions_swithcVSsmooth}. Last two-columns: relic abundance values obtained via FEM.}
\label{tab:inverse_predictions_gammaVsC_smooth}
\end{table}
We address the \textit{inverse} problem of determining $C$ given $\gamma$ that satisfies $\Omega h^2 = 0.120 \pm 0.001$~\cite{Planck:2018vyg}, taking the same approach as Section~\ref{sec:inverseC}. However, we will work in a regime with the transition temperature at $x_t =1$, where no analytical approximation can be obtained. A full numerical treatment via PINNs is then required to obtain physical results. In this regime, the parameterization of the transition between power-law and Hubble law is of paramount importance. We will solve BE~\eqref{eq:freezein2} using the switch transition function~\eqref{eq:Fnew}, as well as the new proposed BE~\eqref{eq:freezein_smooth} using the smooth function~\eqref{eq:smooth_func} with $\beta =2$. As input to the \textit{inverse} PINNs, we take the discrete set of $\gamma$ values presented in Table~\ref{tab:inverse_predictions_gammaVsC_2}. The prior distribution from which we choose the starting value of $C$ is the one provided in Eq.~\eqref{eq:Cprior}. Regarding the hyperparameters of the PINN, the learning rate $\eta_C$ was readjusted to make convergence possible, by lowering its initial value to $1\times 10^{-4}$. The other learning rate, $\eta$, is the same as in Sections~\ref{sec:inverseC} and \ref{sec:inversegamma}, as well as the loss weights.

Our results are gathered in Fig.~\ref{fig:inverse_predictions_swithcVSsmooth} and Tables~\ref{tab:inverse_predictions_gammaVsC_switch} and~\ref{tab:inverse_predictions_gammaVsC_smooth}. From the figure we remark that, compared to the case shown in Fig.~\ref{fig:inverse_predictions_gammaVsC}, the \textit{inverse} PINN found a much different pair-wise relationship between the power-law exponent and interaction cross-sections that satisfies the observed relic abundance. 
We notice that for the switch-like function~(black curves) in the transition temperature regime $z_t =0$, the $C$ values only span over a single order of magnitude, $[-12.2330,-11.1828]$ (see Table~\ref{tab:inverse_predictions_gammaVsC_switch}). When $z_t =\ln 10$, and given the same range in power-law exponents $\gamma \in[-0.7969,2.2922]$, $C$ spanned over seven orders of magnitude, namely $C\in [-14,-7]$. This shows that the new transition temperature regime leads to non-linear effects reflected in the BE. Additionally, we notice that for a given cosmology with negative (positive) exponent, smaller (larger) cross sections are required to reproduce the data in the case of the smooth function (orange curve) when compared to the switch-like transition (black curve) (see Table~\ref{tab:inverse_predictions_gammaVsC_smooth}). Here, we also use FEM as a cross-check of the \textit{inverse} PINN results, with the last column in both tables showing overall agreement in the DM relic abundance prediction of both methods up to the fourth decimal place. This is beyond the accuracy requirement needed in reproducing the experimental data point $\Omega h^2 = 0.120 \pm 0.001$~\cite{Planck:2018vyg}.

Overall, by proposing the smooth transition we showed, without loss of generality, that the PINN can in principle tackle the BE~\eqref{eq:freezein2} for any given $F(x,x_t,\gamma)$ function. This shows the adaptability of the PINN to solve a potentially large class of freeze-in DM BEs. Evidently, the full resolution of the modified Friedmann equations in such alternative cosmologies together with the FIMP BE would yield the full results. This is beyond the scope of this work, but the treatment of such coupled system of equations using PINNs is an interesting problem we leave open for future studies.

\subsection{Bayesian PINN: a method to quantify epistemic uncertainties}
\label{sec:bayesian}

%
    \begin{figure*}[t!]
        \centering
        \includegraphics[scale=0.3]{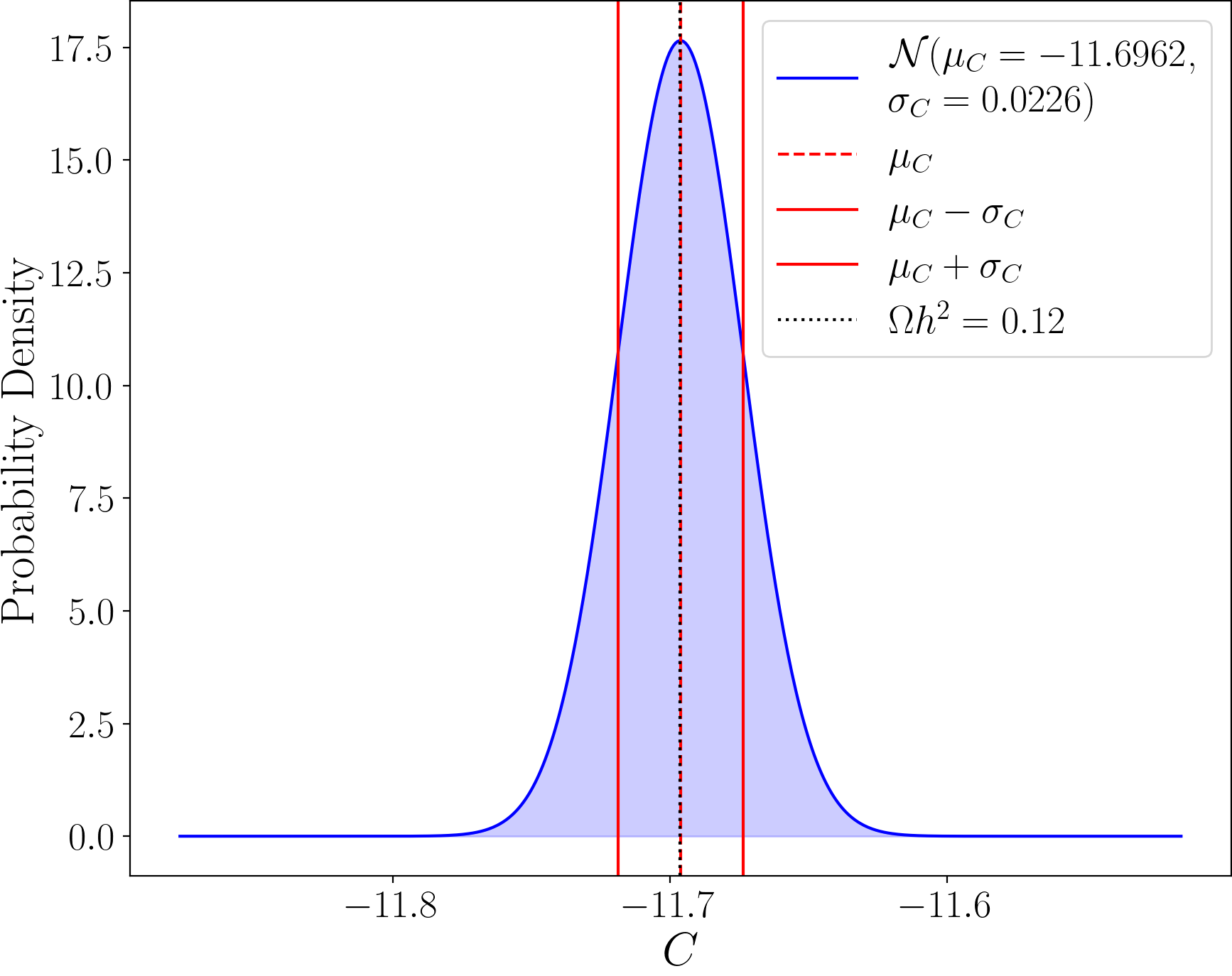} \\
        \includegraphics[scale=0.3]{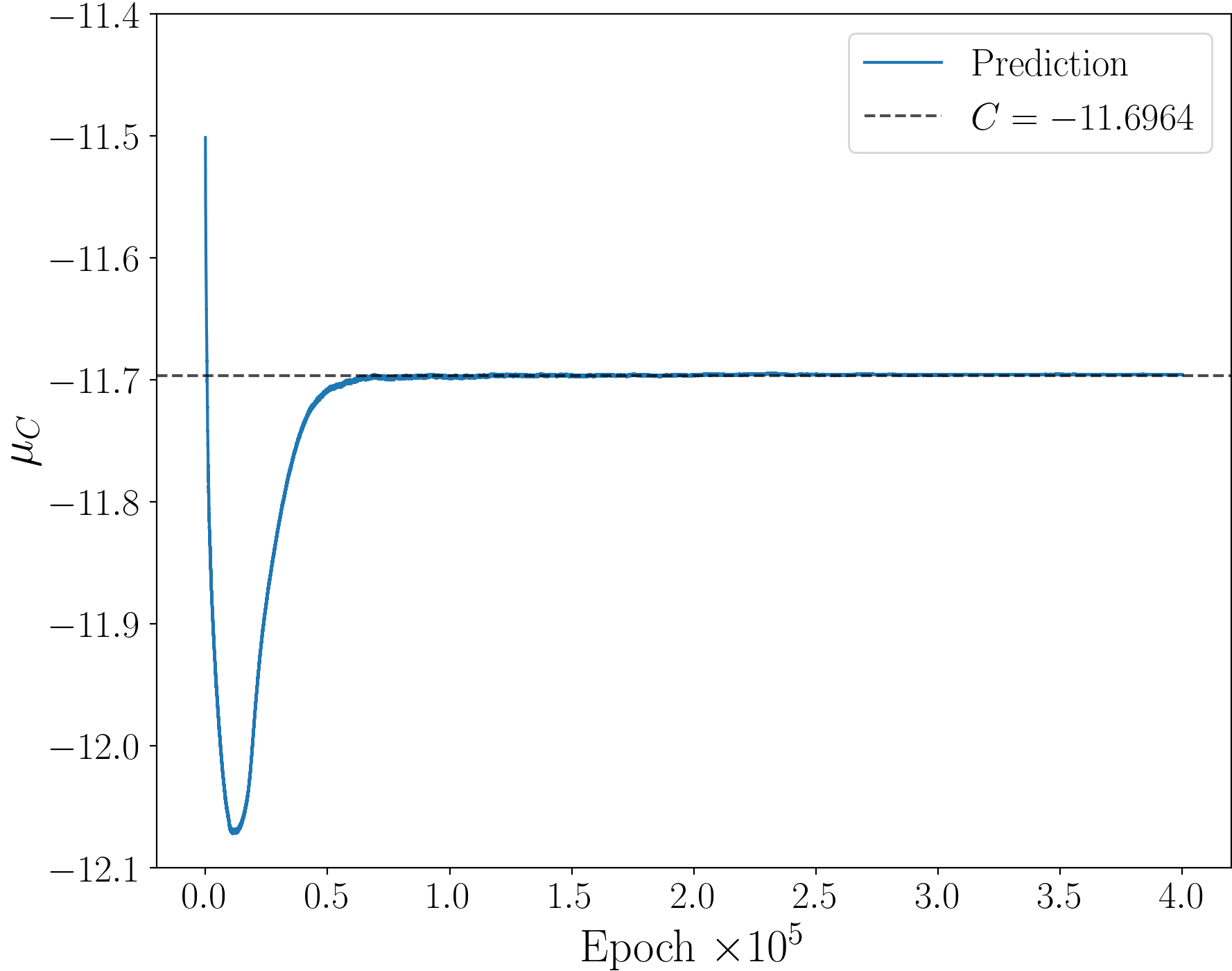} \hspace{+0.1cm} \includegraphics[scale=0.3]{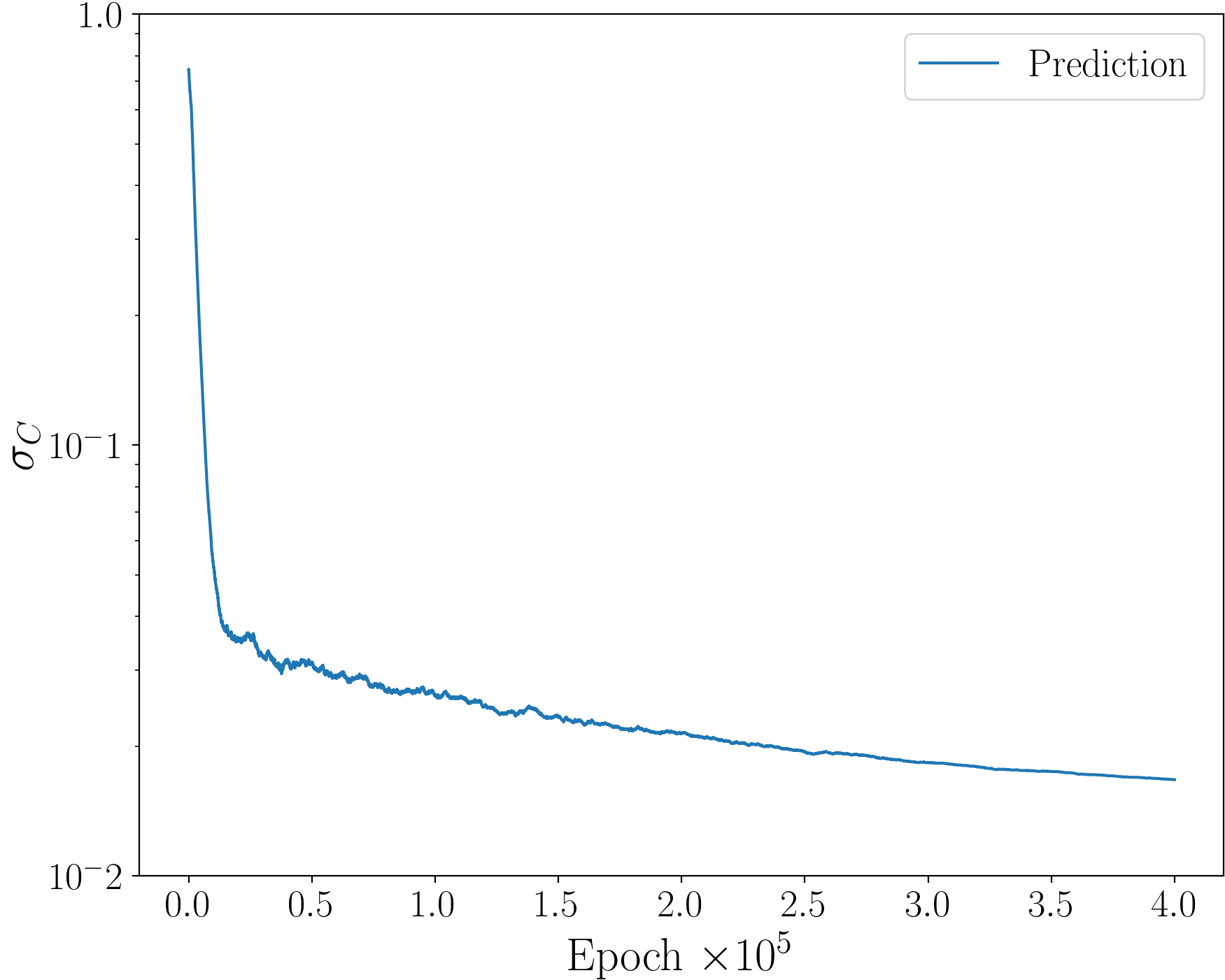}
        \caption{Bayesian PINN results for the \textit{inverse} problem: determining particle interaction strength for Standard cosmology, for the switch transition~\eqref{eq:freezein2} and $z_t = \ln 10$ (see Section~\ref{sec:inverseC}). Gaussian probability density $\mathcal{N}$, the mean $\mu_C$ and standard deviation $\sigma_C$ [see Eq.~\eqref{eq:mucsigmac}], representing the epistemic uncertainty in determining $C$ (see text for details). Evolution of the mean $\mu_C$ (bottom left) and standard deviation $\sigma_C$ (bottom right) in terms of the number of epochs.}
    \label{fig:Bayesian}
    \end{figure*}
Given the nature of an \textit{inverse} problem, i.e. finding a parameter $C$, one could wonder about the possibility of assigning uncertainty to the result of the PINN. Indeed, this is a type of epistemic uncertainty, in which given a prior knowledge of the \textit{inverse} parameter $q(C)$, we measure our lack of knowledge in the posterior $p(C)$.

To enforce the Bayesian technique, we begin by sampling $C$ from a normal distribution -- our prior. Without loss of generality, we choose the prior to be the one in Eq.~\eqref{eq:Cprior}. At each training step, using the reparameterization trick~\cite{NEURIPS2018_92c8c96e, FU2006575, kingma2022autoencodingvariationalbayes},
\begin{equation}
    C_\mathrm{sample} = \mu_C + \sigma_C \cdot \epsilon \; , \quad \epsilon \sim \mathcal{N}(0, 1) \; ,
    \label{eq:mucsigmac}
\end{equation}
a $C$ value was sampled, and the mean $\mu_C$ and standard deviation $\sigma_C$ updated during backpropagation. In accordance with the Bayesian program, we also enforce a regulator in the form of the Kullback–Leibler~(KL) divergence. Thus, the loss function becomes
\begin{equation}
    \mathcal{L}_{\text{Bayesian}} = \mathcal{L}_{\text{Inv}} + \lambda_{\text{KL}}\, D_{\text{KL}}[p(C) 
    \,||\, q(C)] \; ,
\end{equation}
where $\mathcal{L}_{\text{Inv}}$ is given in Eqs.~\eqref{eq:Linv} and~\eqref{eq:LOh2} [see also Eq.~\eqref{eq:Lfwd}], and we chose the weight of $\lambda_{\text{KL}} = 10^{-2}$. The obtained results, summarized in Fig.~\ref{fig:Bayesian}, show the evolution of the training and the final uncertainty. This uncertainty reflects the model's confidence in the found value of $C$. Then, even with a single, noise-free data point, the posterior $p(C)$ does not collapse. The uncertainty, parameterized by~$\sigma_C$, arises because, while $W_{\Omega h^2}$ [see Eq.~\eqref{eq:LOh2}] perfectly constrains this \textit{inverse} solution, the model takes into consideration our prior knowledge, as well as its own inability to perfectly fit a solution. This method, while not describing uncertainty regarding noisy data, or even the model's own weights, is a way to gauge how confident we are in the solution.

Although this uncertainty is parametric, i.e. parameterized by $\lambda_{\text{KL}}$, we found it to be useful in cases where FEMs are not entirely trustworthy. It is also noteworthy that without the KL divergence, the posterior was also found not to collapse entirely. The remaining small amount of uncertainty is then completely related to the optimization process, and was found to be of the order of $10^{-4}$, expected given our previous results in the non-Bayesian \textit{inverse} PINN [see Section~\ref{sec:inverseC}].

\section{Conclusions and outlook}
\label{sec:concl}

ML has revolutionized numerous scientific fields, however its full impact in theoretical particle physics is yet to come. Despite being still in its youth, PINNs have the potential to change this picture, and the main goal in this work was to showcase it. Concretely, we make use of PINNs to solve the BEs governing freeze-in DM in alternative cosmologies.

By addressing the \textit{forward} problem, where is given to the PINN, the BE, particle interactions and cosmologies, we were able to determine the particle relic abundance. The solution for particle yield~$W(z, C)$, is obtained parametrically for cross-section values within an interval, for Standard, RS and GB cosmologies. Compared to FEMs, the \textit{forward} PINN provides a mesh-free method. Indeed, the solution $W(z, C)$ is not computed on a given grid. Instead, a set of collocation points~$z$, is given to the PINN with the objective of minimizing the residuals of the BE, the NN being a continuous function approximating $W(z, C)$. Also, we noticed that the FEM solver is very sensitive to the IC values, sometimes becoming unstable, which does not occur with PINNs.

We showed how PINNs are a particularly powerful technology in what regards \textit{inverse} problems, with FEMs not being inherently designed to solve them. Through \textit{inverse} PINNs, using a singular experimental point, the observed CDM relic density, we determine the physical attributes of the theory: cross-sections and cosmologies. The expansion of the Universe in alternative cosmologies was initially parameterized through a switch-like function characterizing the transition between power and Hubble law. We tested \textit{inverse} PINNs for a new proposed smooth transition and showed, without loss of generality, the adaptability of the PINN to solve a potentially large class of freeze-in DM BEs. The results determined a distinct pair-wise relationship between power-law exponent and particle interactions. Specifically, compared to the switch-function, for a given cosmology with negative (positive) exponent, smaller (larger) cross sections are required to reproduce the data. Finally, via Bayesian methods, we provided a guideline to quantify the epistemic uncertainty of theoretical parameters found in \textit{inverse} problems. 

Although \textit{inverse} PINNs offer a flexible and powerful alternative to traditional FEMs, several limitations remain. The performance of PINNs is highly sensitive to the choice of hyperparameters, that require extensive manual tuning in the absence of a systematic optimization strategy. Furthermore, handling stiffness in differential equations remains a significant challenge. For instance, while the freeze-in BE considered here is not explicitly stiff, the freeze-out regime leads to Riccati-type equations that are explicitly stiff, with sharp transitions in the solution requiring careful numerical treatment. PINN architectures also need to be tailored to the specific form of the underlying ODEs or PDEs, limiting generalizability. Additionally, training PINNs can be computationally expensive, particularly when high accuracy is required or when dealing with complex systems involving multiple coupled equations. Addressing these limitations is essential for scaling PINNs to broader applications in particle physics and cosmology.

In conclusion, PINNs provide a valuable model-building theoretical tool useful in unraveling theories that can explain experimental data. Future directions include the full resolution of coupled systems, such as modified Friedman equations in the context of alternative cosmologies and BE governing DM abundance. Moving towards a complete particle physics model, where we compute the interactions cross-sections, could ultimately make PINNs an interesting alternative to the conventional numerical packages for DM studies. A promising avenue is the development of modular PINNs~\cite{markidis2024braininspiredphysicsinformedneuralnetworks}, in which multiple networks are linked -- for example, one network could approximate the thermally averaged cross section, while another solves the BE to extract the particle abundance. Furthermore, the inclusion of multiple experimental constraints, besides the CDM relic abundance, in a concrete model realization would be a challenging but interesting problem for \textit{inverse} PINNs. Possibilities for further applications of this algorithm in the realms of cosmology, particle and astroparticle physics are endless.

Code is available at \url{https://github.com/MPedraBento/pinn-freeze-in/tree/main}.

\vspace{-0.3cm}

\section*{Acknowledgments}
This research is supported by Fundação para a Ciência e a Tecnologia (FCT, Portugal) through the projects CFTP FCT Unit UIDB/00777/2020 and UIDP/00777/2020, CERN/FIS-PAR/0019/2021 and 2024.02004.CERN, which is partially funded through POCTI (FEDER), COMPETE, QREN and EU. The work of H.B.C. is supported by the PhD FCT grant 2021.06340.BD.




\end{document}